\documentclass{article}

\usepackage{arxiv}

\usepackage[utf8]{inputenc} 
\usepackage[T1]{fontenc}    
\usepackage{hyperref}       
\usepackage{url}            
\usepackage{booktabs}       
\usepackage{amsfonts}       
\usepackage{nicefrac}       
\usepackage{microtype}      

\usepackage{xcolor}
\usepackage{bbding}
\usepackage{enumitem}
\usepackage{amsmath}
\usepackage{amssymb}
\usepackage{subfigure}
\usepackage{color}
\usepackage{graphicx}
\usepackage{algorithm,algpseudocode,algorithmicx}
\usepackage{amsmath,latexsym,amssymb,amsthm,array,amsfonts,algorithm,algpseudocode,booktabs,graphicx,subfigure,multirow,cuted,stfloats}
\usepackage{footmisc}
\usepackage{cite}
\usepackage{units}
\usepackage{tabularx}
\usepackage{multirow}
\usepackage{bm}
\usepackage{float}
\usepackage{soul}
\newlength\savewidth

\author{
  David S.~Hippocampus\thanks{Use footnote for providing further
    information about author (webpage, alternative
    address)---\emph{not} for acknowledging funding agencies.} \\
  Department of Computer Science\\
  Cranberry-Lemon University\\
  Pittsburgh, PA 15213 \\
  \texttt{hippo@cs.cranberry-lemon.edu} \\
   \And
 Elias D.~Striatum \\
  Department of Electrical Engineering\\
  Mount-Sheikh University\\
  Santa Narimana, Levand \\
  \texttt{stariate@ee.mount-sheikh.edu} \\
}

\begin{document}
%
\title{Distributed Additive Encryption and Quantization for Privacy Preserving Federated Deep Learning}
%
%
%

\author{
 Hangyu Zhu\thanks{Equal contribution.}\\
  Department of Computer Science\\
  University of Surrey\\
  Guildford, Surrey, UK \\
  \texttt{hangyu.zhu@surrey.ac.uk} \\
  \And
 Rui Wang \footnotemark[1]\\
  Department of Intelligent Systems\\
  Delft University of Technology\\
  Delft, The Netherlands \\
  \texttt{R.Wang-8@tudelft.nl} \\
   \And
 Yaochu Jin \\
  Department of Computer Science\\
  University of Surrey\\
  Guildford, Surrey, UK \\
  \texttt{Yaochu.Jin@surrey.ac.uk} \\
  \And
  Kaitai Liang \\
  Department of Intelligent Systems\\
  Delft University of Technology\\
  Delft, The Netherlands \\
  \texttt{Kaitai.Liang@tudelft.nl} \\
  \And
  Jianting Ning \\
  School of Information Systems\\
  Singapore Management University\\
  Singapore 178902 \\
  \texttt{jtning88@gmail.com} \\
}

\maketitle

\begin{abstract}

Homomorphic encryption is a very useful gradient protection technique used in privacy preserving federated learning. However, existing encrypted federated learning systems need a trusted third party to generate and distribute key pairs to connected participants, making them unsuited for federated learning and vulnerable to security risks. Moreover, encrypting all model parameters is computationally intensive, especially for large machine learning models such as deep neural networks. In order to mitigate these issues, we develop a practical, computationally efficient encryption based protocol for federated deep learning, where the key pairs are collaboratively generated without the help of a third party. By quantization of the model parameters on the clients and an approximated aggregation on the server, the proposed method avoids encryption and decryption of the entire model. In addition, a threshold based secret sharing technique is designed so that no one can hold the global private key for decryption, while aggregated ciphertexts can be successfully decrypted by a threshold number of clients even if some clients are offline. Our experimental results confirm that the proposed method significantly reduces the communication costs and computational complexity compared to existing encrypted federated learning without compromising the performance and security.
\end{abstract}


%

\section{Introduction}
%
%
%
%
Federated learning (FL) \cite{mcmahan2017communication} enables different clients to collaboratively train a global model by sending local model parameters or gradients to a sever, instead of the raw data. Compared to traditional centralized learning, FL cannot only address the problem of isolated data island, but also play an important role in privacy preservation. Consequently, FL has been deployed in an increasing number of applications in mobile platforms, healthcare, and industrial engineering, among many others \cite{LI2020106854,yang2019federated}.

However, FL consumes a considerable amount of communication resources. Model parameters or gradients need to be downloaded and uploaded frequently between the server and clients in each communication round, resulting in a longer operational stage especially for those large and complex models such as deep convolutional neural networks \cite{krizhevsky2017imagenet}. Much research work has been proposed to reduce the communication costs in the context of FL, including layer-wise asynchronous model update \cite{8945292}, reduction of the model complexity \cite{zhu2020real}, optimal client sampling \cite{ribero2020}, and model quantization \cite{amiri2020federated,xu2020ternary} based on the trained ternary compression \cite{zhu2016trained}. It has been empirically and theoretically shown that using quantization in FL does not cause severe model performance degradation \cite{amiri2020federated,dai2019hyper,du2020high}, and under certain conditions, quantization can even reduce weight divergence \cite{xu2020ternary}.

Although FL can preserve data privacy to a certain degree, recent studies \cite{shokri2015privacy,orekondy2018gradient,8241854,geiping2020inverting,8712695} have shown that local data information can still be breached through the model gradients uploaded from each client.
Under the assumption that the server is \emph{honest-but-curious}, differential privacy (DP) \cite{10.1007/978-3-540-79228-4_1} and homomorphic encryption (HE) \cite{gentry2009fully} have been introduced into FL as additional privacy-preserving mechanisms. DP injects Gaussian \cite{abadi2016deep} or Laplacian \cite{shokri2015privacy} noise into each uploaded model gradients \cite{geyer2017differentially,wei2020federated}, which is cost-effective and light-weighted. But the added noise has a negative impact on the model performance and cannot deal with data reconstruction attacks upon the model gradients \cite{8241854}.

HE is originally applied to outsourced data for privacy-preserving computation and supports direct ciphertext calculations. It can be roughly divided into partial homomorphic, somewhat homomorphic and fully homomorphic encryptions \cite{gentry2009fully}. Among them, additive HE \cite{Paillier1999public} can be seen as a partial HE that provides an efficient protection on gradients and enables gradient aggregation in FL, in which only the addition operation is involved. In an HE-based FL, each client encrypts its model gradients before uploading to the central server, where all encryptions can be directly aggregated. Phong \emph{et al.} \cite{8241854} adopted learning with error (LWE) based HE and Paillier encryption in distributed machine learning to safeguard communicated model parameters. But their method requires that all clients are honest, which is a very strong assumption and unrealistic for many real-world applications. Later, Truex \emph{et al.}\cite{10.1145/3338501.3357370} performed threshold Paillier encryption \cite{damgaard2001generalisation} to construct a more practical system, which is more robust to malicious clients in the FL context.

Similar ideas that use HE in FL have also been presented in \cite{mandal2019privfl,hao2019towards}, which, however, incur a considerable increase in communication costs. More recently, Zhang \emph{et al.} \cite{254465} proposed a batchcrypt scheme for batch gradients encoding and encryption without increasing the communication costs, but this method consumes huge amount of local computational resources.

Using HE and DP \cite{10.1145/3338501.3357370,8825829,8747377} together in FL has become a popular research topic nowadays, which can defend against attacks upon the shared global model on the server side. Nevertheless, DP protection in this framework has limited effect against inference attacks from the client side, since local data has been already ``masked" by HE.

Existing HE-based FL systems have the following two major drawbacks. First, most methods require a trusted third party (TTP) to generate and distribute key pairs, increasing topological complexity and attack surfaces of FL systems. Second, existing HE-based FL designs do not scale well to deep learning models containing a large number of parameters, because encryption and decryption of all trainable parameters are computationally prohibitive and uploading the entire encrypted model consumes a huge amount of communication resources.

To address the above challenges, this work aims to propose a practical and efficient privacy-preserving federated deep learning  framework on the basis of additive ElGamal encryption \cite{ElGamal1985public} and ternary quantization of the local model parameters, DAEQ-FL for short. The main contributions of the work are:
\begin{itemize}
\item We propose, for the first time, an efficient threshold encryption system with federated key generation and model quantization for federated deep learning systems. As a result, the number of model parameters to be encrypted and the communication costs for uploading the local models are considerably reduced, and an extra TTP is no longer required.
\item An approximate model aggregation method is developed for the server to separately aggregate the uploaded ciphertexts and the quantized gradients, making it possible to download the aggregated ciphertexts to $T$ (threshold value) qualified clients only for distributed partial decryption. Thus, the proposed approximate aggregation can further dramatically reduce the computational and communication cost for threshold decryption.
\item Extensive empirical experiments are performed to compare the proposed method to threshold Paillier \cite{10.1145/3338501.3357370} with respect to the learning performance, communication cost, computation time and security. Our results confirm that even encoded with 10 bits, the proposed method can significantly decrease the computation time and communication cost with no or negligible performance degradation.
\end{itemize}

The remainder of the paper is structured as follows. Section II introduces the preliminaries of federated learning, deep neural network models, and the ElGamal encryption system. Details of the proposed methods, including federated key generation, encryption and decryption, ternary quantization, and model aggregation are provided in Section III. Experimental results and discussions are given in Section IV. Finally, conclusion and future work are presented in Section V.
\section{Preliminaries}
\subsection{Federated Learning}
\begin{figure}
\includegraphics[height=4cm, width=7.5cm]{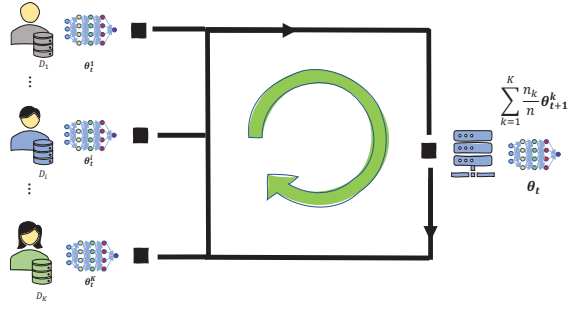}
\centering
\caption{Flowchart of federated learning. $ {\theta_t}  $ are the global model parameters in the $t$-th communication round, ${n_k}$ is the data size of client $k$, and $K$ is the total number of clients. The global model parameters are randomly initialized at the beginning of the training and will be updated by aggregating the uploaded local models in each round.}
\label{hfederated}
\end{figure}

Unlike the centralized cloud model training that needs to collect raw data from different parties and store the data on a server, FL \cite{yang2019federated} is able to distributively learn a shared global model without accessing any private data of the clients. As shown in Fig. \ref{hfederated}, at the $t$-th round of FL, $K$ connected clients download the same global model $\theta_{t}$ from the server and update it by training with their own data. After that, trained local models or gradients will be uploaded back to the server for model aggregation. Therefore, the global model can be learned and updated while all the training data remain on edge devices.

FL aims to optimize a distributed loss function $\ell (\theta )$ as shown in Eq. (\ref{fl}),
\begin{equation}
\begin{split}
\mathop {\min }\limits_\theta \ \ell (\theta ) = \sum\limits_{k = 1}^K {\frac{{{n_k}}}{n}{L_k}(} \theta ), {L_k}(\theta ) = \frac{1}{{{n_k}}}\sum\limits_{i \in {P_k}} {{\ell _i}(\theta, x_i)} \\
\end{split}
\label{fl}
\end{equation}
where $k$ is the index of $K$ total clients, ${L_k}(\theta )$ is the loss function of the $k$-th local client, ${n_k}$ equals to the local data size, and ${P_k}$ is the set of data indexes whose size is $ {n_k} $, i.e., $ {n_k} = |{P_k}|$. Note that the training data $x_i$ on each client $k$ may not satisfy the independent and identically distributed assumption, i.e., non-IID.

Although the data on the clients do not need to be shared, FL is still subject to security risks, since the model gradients naturally contain information of the training data. It has been theoretically proved in \cite{8241854} that only a portion of the gradients may lead to the leakage of private information of the local data. If we assume the cost function to be a quadratic function, the corresponding gradients can be calculated in Eq. (\ref{gradientleak}).
\begin{equation}
\begin{split}
&J(\theta, x)\overset{\text{def}}{=}(h_{\theta}(x)-y)^{2} \\
&g_{k}=\frac{\partial J(\theta, x)}{\partial \theta}=2(h_{\theta}(x)-y)\sigma^{'}(\sum_{i=1}^{d}x_{i}w_{i}+b)\cdot x_{k}
\end{split}
\label{gradientleak}
\end{equation}
where $g_k$ and $x_k$ are the $k$-th feature of gradient $g\in R^{d}$ and the input data $x\in R^{d}$, respectively. Since the product $2(h_{\theta}(x)-y)\sigma^{'}(\sum_{i=1}^{d}x_{i}w_{i}+b)$ is a real scalar number, the gradient is in fact proportional to the input data. As a result, uploading model gradients cannot completely prevent local data from being revealed and enhanced protection techniques are required for secure FL systems.

\subsection{Deep Learning}
Deep learning has been deployed in the fields of computer vision, speech recognition and many other areas \cite{goodfellow2016deep,lecun2015deep,schmidhuber2015deep,deng2014deep,nielsen2015neural}. The word "deep" means that neural network (NN) models, such as convolutional neural networks (CNNs) \cite{lecun1995convolutional} and recurrent neural networks (RNNs) \cite{schuster1997bidirectional}, used in deep learning always contain multiple hidden layers.


For a typical supervised learning \cite{moller1993scaled}, the training purpose is to minimize the expected distance between a desired signal $y$ (e.g., a label in classification) and a predicted value $\hat{y}$, which is often represented by a so-called loss function $\ell (y,\hat{y})$ as shown in Eq. (\ref{loss}), where $\theta$ is the trainable model parameters we want to optimize.
\begin{equation}
\begin{split}
\mathop {\min } \limits_\theta \ \ell (\theta ) = \frac{1}{N}\sum\limits_i {\ell (y,\hat{y}| \theta ,{x_i})}\quad x_i \in \left\{ {{x_1},{x_2}...,{x_N}} \right\}
\label{loss}
\end{split}
\end{equation}

The stochastic gradient descent (SGD) algorithm is the most widely used optimization method that calculates the partial derivatives of the loss function (\ref{loss}) with respect to each model parameter in $\theta$. The model parameters will be updated by subtracting scaled calculated gradients as shown in Eq. (\ref{sgd}),
\begin{equation}
\begin{split}
{g_t} &= {\nabla _\theta }\ell (\theta ,x) \\
{\theta _{t + 1}} &= {\theta _t} - \eta {g_t}
\label{sgd}
\end{split}
\end{equation}
where $\eta$ is the learning rate and ${g_t}$ is the expected gradient over data samples $x$ at the $t$-th iteration. The model update based on SGD in (\ref{sgd}) is repeatedly performed until the model parameters converge.

DNNs often contain a large number of layers and training very deep models on big datasets is extremely time-consuming. Therefore, DNNs will incur excessive communication and computation costs when they are adopted in FL.

\subsection{Homomorphic Encryption and Secret Sharing}
HE is the most widely used data protection technology in secure machine learning as it supports algebraic operations including addition and multiplication on ciphertexts. An encryption method is called partially HE if it supports addition or multiplication operation, and fully HE if it supports an infinite number of addition and multiplication operations.  Without loss of security and correctness, additive HE fulfills that multiple parties encrypt message  $C_i=Enc_{pk}(m_i)$ and decrypt $\sum_{i=1}^{n}m_i=Dec_{sk}(\prod_{i=1}^{n}C_i)$ using public key and secret key, respectively.

Among those well-studied HE techniques, ElGamal\cite{ElGamal1985public} is a multiplicative mechanism while Paillier \cite{Paillier1999public} provides additive operations, which are based on discrete logarithm and composite degree residuosity classes, respectively.
The former needs 256-bit key length to achieve the 128 bit security level, whereas the latter costs 3072 bits \cite{article}, implying that ElGamal is a computationally more efficient encryption and decryption method \cite{Knirsch20a}. However, additional Cramer transformation \cite{cramer1997secure} needs to be applied to ElGamal encryption so as to extend it to support additive operations.

Conventional HE is not well suited for distributed learning systems such as FL systems, since the ciphertexts on the server can be easily inferred, as long as one client uploads its private key to the server. In order to mitigate this issue, Adi Shamir \cite{shamir1979share} proposed Shamir Secret Sharing (SSS) which splits a secret into $n$ different shares. Consequently, $T$-out-of-$n$ shares are needed to recover the secret. Based on SSS and Diffie–Hellman (DH) security definition (Appendix \ref{sedef}), Feldman proposed verifiable secret sharing (VSS) \cite{feldman1987practical}, which adds a verification process during sending shares.


But Feldman VSS only allows trusted clients to share secrets and it will fail to generate correct key pairs if there are adversaries in the system. To address this limit, Perdersen \cite{pedersen1991non} proposed a novel VSS that is able to detect and exclude adversarial clients. On the basis of the above two VSS methods, Gennaro \emph{et al.} \cite{gennaro1999secure} introduced a secure distributed key generation (DKG) for discrete logarithm based crytosystems.

We propose a federated key generation (FKG) based on DKG for model parameter encryption in FL and the details of FKG will be given in Section III.

\section{Our Proposed System}

A key component for the proposed system is federated key generation (FKG) that generates keys for all participating clients of the FL system without a TTP. To this end, additive discrete logarithm based encryption is adopted to achieve secure model aggregation. In addition, a fixed point encoding method is implemented to encode the plaintext. In order to decrease computational and communication resources, ternary gradient quantization and approximate model aggregation are further introduced. In the following, we elaborate each of our main components and present a description of the proposed overall DAEQ-FL system. Finally, a brief discussion is given to compare our DAEQ-FL with existing encryption based FL systems.

\subsection{Federated Key Generation}
The proposed FKG is a variant of DKG \cite{gennaro1999secure}, which is based on Pedersen VSS and Feldman VSS that can verify key shares in a secure way for successful key generations.
The main steps of FKG are described in \textbf{Algorithm \ref{fkg}}.

\begin{algorithm}[!htb]\footnotesize{
\caption{Federated Key Generation. $\mathbb{G}$ is a cyclic group, $p$ and $q$ are large prime numbers, $g$ is a generator, $y \in \mathbb{G}$, $N$ is the number of total clients, $C$ is the fraction of clients participating the current round, $i$ is the client index, and $T$ is the threshold value.}
\algblock{Begin}{End}
\label{fkg}
\begin{algorithmic}[1]
\State \textbf{\emph{Server distributes public parameters $<p, q, g, y>$}}
\For{Each FL round $ t = 1,2,...\ $}
\State $n = C*N$
\State Select threshold value $T>n/2$
\State {Client $ i \in \{1,\cdots,n\} $ perform \textbf{Pedersen VSS}}
\State Collect the number of complaints $cpt_i$ for client $i$
\For {Client $ i \in \{1,\cdots,n\}$}
\If{$cpt_i > T$}

\State Mark client $i$ as disqualified
\Else
\State Client $i$ uploads $f_{i}(j)$ and
\If{Eq. (\ref{pedersencommit}) is satisfied}
\State Mark client $i$ as qualified (QUAL)
\Else
\State Mark client $i$ as disqualified
\EndIf
\State Mark client $i$ as QUAL \Comment{$T\leq$ |QUAL| $\leq n$}
\EndIf
\EndFor
\State {Client $ i \in$ QUAL perform \textbf{Feldman VSS}}
\State Collect complained client index in $O$,
\For{Each client $i\in$ O} \Comment{$|O|<T$}
\State Set $counter=0$
\For{Each client $j\in$ QUAL but $j\notin m$}
\State Client $j$ uploads $f_{j}(i)$ and $f_{j}^{'}(i)$
\If{Eq. (\ref{feldmancommit}) is satisfied}
\State $counter=counter+1$
\EndIf
\If{$counter\geq T$}
\State Break
\EndIf
\EndFor
\State Retrieve $f_{i}(z)$  and $A_{i0}$ \Comment{$A_{i0}=g^{a_{i0}}\pmod{p}$}
\EndFor
\State Generate global public key $h=\prod_{i \in QUAL}$ $A_{i0}=g^x$
\EndFor
\end{algorithmic}}
\end{algorithm}

Assume that the server in DAEQ-FL is \emph{honest-but-curious}, and there are at least $T$-out-of-$n$ ($T> n/2$) honest clients. Before key pair generation, the server needs to generate and distribute four public parameters $p$, $q$, $g$ and $y$, where $q$ is the prime order of cyclic group $\mathbb{G}$, $p$ is a large prime number satisfying $p-1=rq$, $r$ is a positive integer, $g$ and $y$ are two different random elements in $\mathbb{G}$.

For Pedersen VSS executed in line 5 of \textbf{Algorithm \ref{fkg}}, each participating client $i$ in the $t$-th round generates two random polynomials $f_{i}(z)$ and $f_{i}^{'}(z)$ over $\mathbb{Z}_{q}^{*}$ of order $T-1$ as shown in Eq. (\ref{fkgpvss}).
\begin{equation}
\begin{split}
f_{i}(z) &= a_{i0} + a_{i1}x\,+...+\,a_{iT
-1}x^{T-1}\pmod{q} \\
f_{i}^{'}(z) &= b_{i0} + b_{i1}x\,+...+\,b_{iT-1}x^{T-1}\pmod{q}
\end{split}
\label{fkgpvss}
\end{equation}
Let $z_{i}=a_{i0}=f_{i}(0)$ be the locally stored private key. Client $i$ broadcasts $C_{ik}=g^{a_{ik}}y^{b_{ik}}\pmod p$ and sends shares $s_{ij}=f_{i}(j)$, $s_{ij}^{'}=f_{i}^{'}(j)$ to client $j$ ($j\in n$), then client $j$ verifies the received shares by \emph{Pedersen commitment}:
\begin{equation}
\begin{split}
g^{s_{ij}}y^{s_{ij}^{'}}=\prod_{k=0}^{T-1}(C_{ik})^{j^{k}} \pmod{p}
\end{split}
\label{pedersencommit}
\end{equation}


Due to the hiding and binding properties of Pedersen commitment \cite{pedersen1991non}, it is impossible for adversaries, if any, to guess the real $a_{ik}$ and $b_{ik}$ through $C_ik$ or to find another a pair of $s_{ij}$ and $s_{ij}^{'}$ that can satisfy Eq. (\ref{pedersencommit}). In addition, based on our previous security assumption and the principle of SSS, it is infeasible to reconstruct the private keys of any honest clients even if the system contains $n-T$ malicious clients.

Each client sends a complaint of client $i$ to the server if any shares $s_{ij}$ and $s_{ij}^{'}$ received from client $i$ do not satisfy Eq. (\ref{pedersencommit}). Once the server receives more than $T$ complaints against client $i$ (line 6 in \textbf{Algorithm \ref{fkg}}), this client will be immediately disqualified. Besides, as long as client $i$ is complained by any client $j$, where $ j \in \{1,\cdots,n\} $, the corresponding shares $s_{ij}$ and $s_{ij}^{'}$ are required to upload to the central server for Pedersen commitment (Eq. (\ref{pedersencommit})) verification. If any verification fails, client $i$ would be marked as disqualified.

However, Pedersen VSS cannot guarantee correct global public key generation, since malicious clients can still corrupt the generation process by broadcasting fake $A_{i0}$ (line 33 in \textbf{Algorithm \ref{fkg}}). Therefore, Feldman VSS is used in addition to Pedersen VSS to ensure that all the QUAL clients broadcast correct $A_{i0}$ for in the proposed FKG.


Similarly, to implement Feldman VSS (line 20 in \textbf{Algorithm \ref{fkg}}), each client $j$ ($j\in$ QUAL) broadcasts $A_{ik}=g^{a_{ik}}\pmod{p}$ and verifies Eq. (\ref{feldmancommit}).
\begin{equation}
\begin{split}
g^{s_{ij}}=\prod_{k=0}^{T-1}(A_{ik})^{j^{k}} \pmod{p}
\end{split}
\label{feldmancommit}
\end{equation}
If shares of client $i$ satisfy Eq. (\ref{pedersencommit}) but not Eq. (\ref{feldmancommit}), client $j$ will send a complaint to server. Then the server requires $t$ QUAL clients to upload their shares $f_{i}(j), j\in t$ to retrieve the random polynomial $f_{i}(z)$ of client $i$ by Lagrange interpolation function \cite{berrut2004barycentric} as show in Eq. (\ref{lagrange}).

\begin{equation}
\begin{split}
&\lambda_{j}=\prod_{k\neq j}\frac{z-k}{j-k},k\in T,j\in \text{QUAL} \\
&f_{i}(z)=\sum_{j\in \text{QUAL}}\lambda_{j}f_{i}(j)
\end{split}
\label{lagrange}
\end{equation}
Finally, the server can generate the global public key through broadcasting $A_{i0}$ from all QUAL clients in Eq. (\ref{pk}), where $x$ is in fact the global private key. And then, the public key $h$ will be shared to all QUAL clients.
\begin{equation}
\begin{split}
&h_{i}=A_{i0}=g^{z_{i}}\pmod{p} \\
&h=\prod_{i\in \text{QUAL}}h_{i}=g^{\sum_{i\in \text{QUAL}}z_{i}}=g^{x}\pmod{p}
\end{split}
\label{pk}
\end{equation}

\subsection{Additive Discrete Logarithm Based Encryption}
To be fully compatible with FKG, which is adapted from DKG to the FL environment, additive discrete logarithm based encryption is employed based on ElGamal encryption \cite{ElGamal1985public}.

The original ElGamal encryption works as follows:
\begin{itemize}
\item Parameters generation: Generate three parameters $p$, $q$ and $g$, where $q$ is the prime order of a cyclic group $\mathbb{G}$, $p$ is a large prime number satisfying $q|p-1$, and $g$ is a generator of $\mathbb{G}$.

\item Key generation: Select a random number $x, x\in \mathbb{Z}_{q}$ as the secret key, and then compute $h=g^{x}\pmod{p}$ to be the public key.

\item Encryption: To encrypt a message $m\in \mathbb{Z}_{p}^{*}$, choose a random number $r\in \mathbb{Z}_{q}^{*}$ as a ephemeral key, calculate two ciphertexts as $<c_{1}=g^{r}\pmod{p}, c_{2}=mh^{r}\pmod{p}>$.

\item Decryption: The plaintext message $m$ can only be decrypted if the private key $x$ is available by computing Eq. (\ref{ElGamaldec})
\begin{equation}
\begin{split}
\frac{c_{2}}{c_{1}^{x}}=\frac{mh^{r}}{(g^{r})^{x}}=\frac{mg^{xr}}{g^{rx}}\pmod{p}\equiv m
\end{split}
\label{ElGamaldec}
\end{equation}
\end{itemize}

Therefore, the original ElGamal is a multiplicative HE satisfying: $\text{Enc}(m1)*\text{Enc}(m2)=\text{Enc}(m1*m2)$, because $m_{1}h^{r_{1}}*m_{2}h^{r_2}=m_{1}m_{2}h^{r_{1}+r_{2}}\pmod{p}$. Since model aggregation on the server in FL performs the addition operation,  we can apply Cramer transformation \cite{cramer1997secure} on ElGamal encryption by simply converting the plaintext $m$ into $m^{'}=g^{m}\pmod{p}$. Consequently, the original ElGamal encryption becomes a discrete logarithm based additive HE, as shown in Eq. (\ref{addElGamal}).
\begin{equation}
\begin{split}
\text{Enc}(m_{1})*\text{Enc}(m_{2})&=g^{m_{1}}h^{r_{1}}*g^{m_{2}}h^{r_{2}} \\
&=g^{m_{1}+m_{2}}h^{r_{1}+r_{2}}\pmod{p}
\end{split}
\label{addElGamal}
\end{equation}

Note that the specific security analysis is described in Appendix \ref{seana}.

\subsection{Fixed Point Encoding Method}
Note that HE can be applied to integers only, however, model parameters or gradients are normally real numbers. Therefore, the real-values model parameters must be encoded before encryption.

The encoding method used in this work is straightforward, as shown in Eq. (\ref{encode}), where $grad$ is a real number gradient, $b$ is the encoding bit length, $q$ is the above mentioned prime order of $\mathbb{G}$ and $m$ is the encoded integer number.
\begin{equation}
\begin{split}
& int_{\text{max}} = \text{int}(q/3) \\
& \widehat{m}=\text{round}(grad*2^{b}),  \widehat{m} \leq int_{\text{max}} \\
& \text{Encode}(grad)=\widehat{m}\pmod{q}=m \\
& \text{Decode}(m)=\left\{\begin{matrix}
& m*2^{-b}, & m\leq int_{\text{max}} \\
& (m-q)*2^{-b}, & m>q-int_{\text{max}}
\end{matrix}\right.
\end{split}
\label{encode}
\end{equation}
$int_{\text{max}}$ is the maximum positive encoding number defined by the server, which is one-third of $q$. If $m>q-int_{\text{max}}$ ($\widehat{m}$ is negative), it should subtract $q$ before multiplying $2^{-b}$, because $\widehat{m}\pmod{q}=q+\widehat{m}, \widehat{m}<0$. Since the bit length of $int_{\text{max}}$ is always set to be much larger than the encoding bit $b$, sufficient value space can be reserved for encoding number summations (additive HE).

\subsection{Brute Force and Log Recovery}
Using Cramer transformation needs to recover the desired $m$ from $m^{'}$ to solve the so-called discrete logarithm hard problem after decryption. Here, we propose two techniques, namely brute force and log recovery, to solve this problem.

Brute force recovery simply tries different $m$ from $0$ to $q-1$, and the correct $m$ is found only if $m^{'}=g^{m}\pmod{p}$. Thus, a maximum of $q$ trials are needed to solve DLHP in the worst case. Fortunately, the absolute values of all the model gradients in DNNs are always less than $0.1$, and therefore, they can be encoded with a $b$ bit-length fixed point integer number. It takes about $2^{b}$ times to find the correct $m$ if $\widehat{m}$ is positive. Note that the quantization method we employ can guarantee that $\widehat{m}$ is positive, which will be introduced later.

The log method consumes almost no additional recovery time by calculating $\text{log}_{g} ({g}^{m})=m$ directly. However, this only works when $g^{m}<p$. In order to ensure the best encoding precision (the encoded $m$ should be as large as possible), we minimize $g$ to $g_{0}=2$. And the encoded number $m$ must be less than the bit length of a large prime number $p$, which means the encoding precision is restricted by the security level. Besides, the security level will not be reduced by selecting a small fixed $g_{0}$.

Both the brute force and log recovery are described in \textbf{Algorithm \ref{recovery}}
\begin{algorithm}[htbp]\footnotesize{
\caption{Plaintext Recovery. $q$ is a prime order of the cyclic group $\mathbb{G}$, $p$ is a large prime number satisfying $p-1|q$, $g$ is a random element in $\mathbb{G}$, $m$ is the message to be recovered, $int_{\text{max}}$ is the maximum positive encoded number.}
\algblock{Begin}{End}
\label{recovery}
\begin{algorithmic}[1]
\State \textbf{Brute Force Recovery: }
\State $g_{0}=g$
\State Given decrypted plaintext $m^{'}=g_{0}^{m}\pmod{p}$, $m\leq int_{\text{max}}$
\For{$j$ from $0$ to $q-1$}
\If{$g_{0}^{j}\pmod{p}==m^{'}$}
\State $m=j$
\State Break
\Else
\State Continue
\EndIf
\EndFor
\State \textbf{Return} $m$ \\
\State \textbf{Log Recovery: }
\State $g_{0}=2$
\State Given decrypted plaintext $m^{'}=g_{0}^{m}\pmod{p}$, $g_{0}^{m}\leq p$
\State $m=\text{log}_{g_{0}}(g_{0}^{m})$
\State \textbf{Return} $m$
\end{algorithmic}}
\end{algorithm}

\subsection{Ternary Gradient Quantization}
Encrypting and decrypting all elements of the model gradients have several shortcomings. First, performing encryption on local clients is computationally extremely expensive, causing a big barrier for real world applications, since usually the distributed edge devices do not have abundant computational resources. Second, uploading model gradients in terms of ciphertext incurs a large amount of communication costs. Finally, the first two issues will become computationally prohibitive when the model is large and complex, e.g., DNNs.

In order to tackle the above challenges, we introduce ternary gradient quantization (TernGrad) \cite{wen2017terngrad} to drastically reduce computational and communication costs for encryption of DNNs. TernGrad compresses the original model gradients into ternary precision gradients with values $\in\{ -1,0,1 \}$ as described in Eq. (\ref{terngard}),
\begin{equation}
\begin{split}
&\tilde{g_{t}}=s_{t}\cdot \text{sign}(g_{t})\cdot b_{t} \\
&s_{t}=\text{max}(\text{abs}(g_{t}))
\end{split}
\label{terngard}
\end{equation}
where $g_{t}$ is the full precision model gradients at the $t$-th iteration, $\tilde{g_{t}}$ is the quantized gradients, $s_{t}$ (a scalar larger than $0$) is the maximum absolute element among $g_{t}$, and the $sign$ function transfers $g_{t}$ into binary precision with values $\in\{ -1,1 \}$. Finally, $b_{t}$ is a binary tensor whose elements follow the Bernoulli distribution \cite{uspensky1937introduction}.
\begin{equation}
\begin{split}
&Pr(b_{tk}=1|g_{t})=|g_{tk}|/s_{t}\\
&Pr(b_{tk}=0|g_{t})=1-|g_{tk}|/s_{t}
\end{split}
\label{bt}
\end{equation}
where $b_{tk}$ and $g_{tk}$ is the $k$-th element of $b_{t}$ and $g_{t}$, respectively, and the product of $\text{sign}(g_{t})$ and $b_{tk}$ is a ternary tensor ($\{ -1,0,1 \}$) representing the model training direction. According to Eq. (\ref{sgd}), the full precision model parameters in TernGrad is updated as shown in Eq. (\ref{terngradsgd}).
\begin{equation}
\begin{split}
\theta_{t+1}=\theta_{t}-\eta (s_{t}\cdot \text{sign}(g_{t})\cdot b_{t})
\end{split}
\label{terngradsgd}
\end{equation}
Since $s_{t}$ is a random variable depending on input $x_{t}$ and the model weights $\theta_{t}$, Eq. (\ref{bt}) can be re-written into Eq. (\ref{bt1}).
\begin{equation}
\begin{split}
&Pr(b_{tk}=1|x_{t},\theta_{t})=|g_{tk}|/s_{t}\\
&Pr(b_{tk}=0|x_{t},\theta_{t})=1-|g_{tk}|/s_{t}
\end{split}
\label{bt1}
\end{equation}
The unbiasedness of the ternary gradients can be proved as shown in Eq. (\ref{unbias}).
\begin{equation}
\begin{split}
\mathbb{E}(s_{t}\cdot \text{sign}(g_{t})\cdot b_{t})&=\mathbb{E}(s_{t}\cdot \text{sign}(g_{t})\cdot \mathbb{E}(b_{t}|x_{t})) \\
&=\mathbb{E}(g_{t})
\end{split}
\label{unbias}
\end{equation}

The TernGrad algorithm is adopted in the proposed DAEQ-FL to significantly reduce the computational cost for encryption on local devices and the communication costs for passing the encrypted model parameters between the clients and the server. Before performing encryption on the local clients, the model gradients are decomposed into two parts: one is the positive scalar $s_{t}$ and the other is the ternary model gradients $\text{sign}(g_{t})\cdot b_{t})$. And only one scalar $s_{t}$ in each layer of the model needs to be encrypted, thereby dramatically decreasing the computation costs and encryption time.

The ternary gradients do not need to be encrypted, because adversaries, if any, can only derive parts of gradients sign information from them. And separately uploading encrypted $s_{t}$ and ternary gradients makes the total amount of uploads 16 times smaller than the original scenarios. Another advantage is that it can make brute force recovery much faster, since the scalar $s_{t}$ can never be negative, namely, the encoded $m$ in Eq. (\ref{encode}).

\subsection{Approximate Model Aggregation} \label{appmodelagg}
For model aggregation on the server, the received encrypted $s_{t}$ ($t$-th round in FL) should multiply its corresponding ternary gradients before weighted averaging (Eq. (\ref{agg})).
\begin{equation}
\begin{split}
g_{(t,\text{tern})}^{i}&=\text{sign}(g_{t}^{i})\cdot b_{t}^{i} \\
\text{Enc}(g_{t}^{\text{global}})&=\sum_{i}\frac{n_{i}}{n}(\text{Enc}(s_{t}^{i})*g_{(t,\text{tern})}^{i})
\end{split}
\label{agg}
\end{equation}

\begin{figure}
\includegraphics[height=4cm, width=8cm]{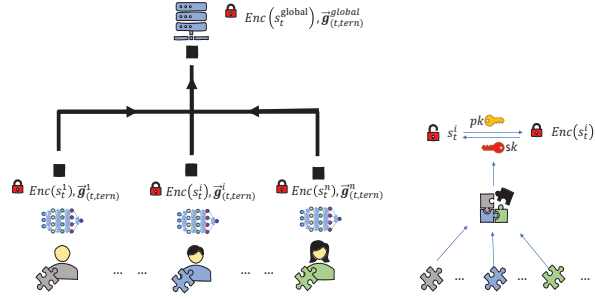}
\centering
\caption{Encryption with TernGrad and model aggregation approximation. The $\text{Enc}(s_{t}^{\text{global}})$ is aggregated over the uploaded $\text{Enc}(s_{t}^{i})$ from the participating clients.}
\label{aggappfigure}
\end{figure}

Decryption of the aggregated global gradients $\text{Enc}(g_{t}^{\text{global}})$ requires to traverse each single ciphertext, which is extremely time-consuming, making it less practical even if the server often possess abundant computation resources. Therefore, this work proposes an approximate aggregation method (Fig. \ref{aggappfigure}) so as to reduce the decryption time. The basic idea is to separately aggregate the encrypted scalar and related ternary gradients, as shown in Eq. (\ref{aggapp}).
\begin{equation}
\begin{split}
\text{Enc}(s_{t}^{\text{global}})&=\prod_{i}\text{Enc}(s_{t}^{i}*\frac{n_{i}}{n})=\text{Enc}(\sum_{i}s_{t}^{i}*\frac{n_{i}}{n}) \\
g_{(t,\text{tern})}^{\text{global}}&=\sum_{i}g_{(t,\text{tern})}^{i} \\
\end{split}
\label{aggapp}
\end{equation}

where $i$ is the client index, $n_{i}$ is the local data size and $n$ is the global data size. And $\text{Enc}(g_{t}^{\text{global}})=\text{Enc}(s_{t}^{\text{global}})*g_{(t,\text{tern})}^{\text{global}}$, if $s_{t}^{1}=s_{t}^{2}= ... =s_{t}^{n}$. However, this condition is hard to satisfy and the bias $g_{t}^{\text{global}}-s_{t}^{\text{global}}*g_{(t,\text{tern})}^{\text{global}}$ is difficult to estimate due to the random property of SGD. In reality, each client's local scalar satisfies $s_{t}^{1} \approx s_{t}^{2} \approx ... \approx s_{t}^{n}$, and the our experimental results also empirically confirm that this approximation bias is acceptable and even negligible.

Note that a small implementation trick used here is to do weighted averaging upon $s_{t}$ before encryption, which helps reduce the differences between $s_{t}^{i}$ and avoid overflow of the encrypted gradients during model aggregations on the server.

To update the global model, only one ciphertext $\text{Enc}(s_{t}^{\text{global}})$ of each layer needs to be decrypted, which will be multiplied by the global ternary gradients afterwards as shown in Eq. (\ref{decrypt})
\begin{equation}
\begin{split}
&s_{t}^{\text{global}}=\text{Dec}(\text{Enc}(s_{t}^{\text{global}})) \\
&\theta_{t+1}^{\text{global}}=\theta_{t}^{\text{global}}-\eta s_{t}^{\text{global}}*g_{(t,\text{tern})}^{\text{global}}
\end{split}
\label{decrypt}
\end{equation}

\subsection{Overall Framework: DAEQ-FL}
The overall framework, distributed additive ElGamal encryption and quantization for privacy-preserving federated deep learning, DAEQ-FL for short, is depicted in \textbf{Algorithm \ref{DAEQ-FL}}. Note that in DAEQ, federated key generation should be performed at the beginning of each communication round before model training to generate different key pairs for preventing possible collusion between a client and the server.

\begin{algorithm}[htbp]\footnotesize{
\caption{DAEQ-FL. $p$, $q$, $g$, $y$ are key parameters introduced in \textbf{Algorithm \ref{fkg}}, $pk$ is the global public key, $Qual$ are qualified clients, $N$ is the total number of clients, $C$ is the fraction of connected clients, $E$ is the number of local epochs, $B$ is the local batch data, $\theta_{t}$ is the global model parameters at the $t$-th FL round, $T$ is the threshold value, and $\eta$ is the learning rate.}
\algblock{Begin}{End}
\label{DAEQ-FL}
\begin{algorithmic}[1]
\State \textbf{Server: }
\State  Generate and distribute $p$, $q$, $g$, $y$ and global model parameters ${\theta_{0}}$
\For {each FL round $ t = 1,2,...\ $}
\State Select $ n = C \times N\ $ clients, $ C \in (0,1)\ $
\State Select $T>n/2$
\State Generate $pk$ by FKG among $n$ clients in \textbf{Algorithm \ref{fkg}}
\For {each client $ i \in Qual$ in parallel}
\State Download $\theta_{t}$
\State Do local \textbf{Training}
\State Upload $c_{(t,1)}^{i}$, $c_{(t,2)}^{i}$ and $\Delta \theta_{(t,tern)}^{i}$
\EndFor
\State $c_{(t,1)}=\prod_{i}c_{(t,1)}^{i}\pmod{p}$
\State $c_{(t,2)}=\prod_{i}c_{(t,2)}^{i}\pmod{p}$
\State $\Delta \theta_{(t,tern)}=\sum_{i}\Delta \theta_{(t,tern)}^{i}$
\State Randomly select $T$ $Qual$ clients
\For{each client $j\in T$ in parallel}
\State Download $c_{(t,1)}$ and $c_{(t,2)}$
\State Do \textbf{\emph{Partial Decryption}}
\EndFor
\State $g_{0}^{Tm_{t}}=\prod_{j\in T}pd_{j}=g_{0}^{Tm_t}g^{(x-\sum_{j}\lambda_{j}x_i)T}\pmod{p}$
\State Recover $Tm_{t}$ by \textbf{Algorithm \ref{recovery}}
\State $\theta_{t+1}=\theta_{t}- \Delta \theta_{(t,\text{tern})}*Tm_{t}/T$
\EndFor \\
\State \textbf{Client $i$: }
\State \textbf{// Training:}
\State $\theta_{t}^{i} = \theta_{t}$
\For {each iteration from 1 to $E$}
\For {batch $ b \in B\ $}
\State $ {\theta_{t}^i} = {\theta_{t} ^i} - \eta \nabla {L_i}({\theta_{t} ^i},b)\ $
\EndFor
\EndFor
\State $\Delta \theta_{t}^{i}=\theta_{t}^{i}-\theta_{t}$
\State Quantize $\Delta \theta_{t}^{i}$ into $s_{t}^{i}$ and $\Delta \theta_{(t,\text{tern})}^{i}$ in Eq. (\ref{terngard}) and (\ref{agg})
\State Encode $m_{t}^{i}=\text{round}(s_{t}^{i}*D_{k}/D*2^{l})\pmod{q}$ in Eq. (\ref{encode})
\State Encrypt $c_{(t,1)}^{i}=g^{r_{i}}\pmod{p}$, $c_{(t,2)}^{i}=g_{0}^{m_{t}^{i}}pk^{r_{i}}\pmod{p}$
\State \textbf{Return} $c_{(t,1)}^{i}$, $c_{(t,2)}^{i}$ and $\Delta \theta_{(t,tern)}^{i}$ to server
\State \textbf{// Partial Decryption: }
\State $x_{i}=\sum_{j}s_{ji}=f_{j}(i),\,i,j\in \text{QUAL}$ \Comment{$f_{j}(i)$ in \textbf{Algorithm \ref{fkg}}}
\State Partial decrypt $pd_{i}=c_{(t,2)}/c_{(t,1)}^{\lambda_{i}x_{i}T}\pmod{p}, r_{i}\in \mathbb{Z}_{q}^{*}$
\State \textbf{Return} $pd_{i}$
\end{algorithmic}}
\end{algorithm}

Note that the server can determine the threshold value $T$ based on the number of participating clients $n$ in each FL round ($T>n/2$). If the number of QUAL clients are less than $T$, the disqualified clients will be kicked out of the system and then the process is aborted and FKG is restarted. After FKG, each QUAL client $i$ downloads the global model parameters $\theta_{t}$ and the public key $pk$ for local training. Then the model gradients, obtained by subtracting the received global model $\theta_{t}$ from the local updated model $\theta_{t}^{i}$, are converted into a real-valued coefficient $s^{i}_{t}$ and a ternary matrix $\Delta \theta_{(t,\text{tern})}^{i}$ before performing ElGamal encryption. Two ciphertexts $c^{i}_{(t,1)}$ and $c^{i}_{(t,2)}$ together with a ternary gradient $\Delta \theta_{(t,\text{tern})}^{i}$ are then uploaded to the server for model aggregation as described in Section \ref{appmodelagg}.

\begin{figure}
\includegraphics[height=4cm, width=8cm]{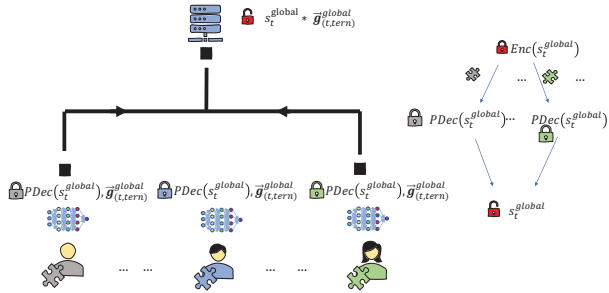}
\centering
\caption{The encrypted $s_{t}^{\text{global}}$ is downloaded to $T$ qualified clients for partial decryption, and then the partial decrypted ciphertexts are uploaded back to the server to retrieve the final plaintext.}
\label{partialdecfigure}
\end{figure}

For decryption (Fig. \ref{partialdecfigure}), only two aggregated ciphertexts need to be downloaded to $T$ QUAL clients for partial decryption and $T$ partial decrypted ciphertexts $pd_{i}$ are uploaded back to the server (line 16-20 in \textbf{Algorithm \ref{DAEQ-FL}}). The server can easily get the plaintext $g_{0}^{Tm_{t}}$ by multiplying all received ciphertexts $pd_{i}$.
$s_{ji}$ is a share $f_{j}(i)$ (introduced in \textbf{Algorithm \ref{fkg}}) from client $j$ to client $i$, $\lambda_{i}=\prod_{j\neq i}\frac{j}{j-i}$ is the Lagrange coefficient and $x=\sum_{j\in \text{QUAL}}z_{j}$ is the global private key. According to the property of SSS, at least $T$ (threshold value) different $f_{i}(j), j\in T$ shares are needed to retrieve the local private key $z_{i}$. The reason why $x-\sum_{i}\lambda_{i}x_i=0$ is proved below:
\begin{equation}
\begin{split}
\sum_{i\in T}\lambda_{i}x_i&=\sum_{i\in T}\lambda_{i}\sum_{j\in \text{QUAL}}f_{j}(i) \\
&=\sum_{j\in \text{QUAL}}\sum_{i\in T}\lambda_{i}f_{j}(i) \\
&=\sum_{j\in \text{QUAL}}z_{j}=x
\end{split}
\label{zero}
\end{equation}

One of the advantages of DAEQ-FL is that no parties within the FL system, including the server, can know the global private key $x$, which significantly enhances system security level. In addition, the extra communication resources are negligible, and only three ciphertexts $c_{(t,1)}$, $c_{(t,2)}$ and $pd_{i}$ are transmitted between the server and each client $i$ with the help of the TernGrad algorithm. Finally, DAEQ-FL is robust to possible disconnection of individual clients, since the ciphertext can be successfully decrypted so long as a minimum of $T$ QUAL clients upload their $pd_{i}$.

\subsection{Discussion}
We list the general differences between our proposed system and four popular existing approaches in Table \ref{comparison}. From the table we can see that our DAEQ-FL has remarkable superiority, being a threshold based encryption system without an extra TTP and tolerating the existence of malicious clients. We here note that Truex \emph{et al.} \cite{truex2019hybrid} also proposed a threshold based Paillier encryption system in FL, but it still requires a TTP for key generation. Some quantization technique is introduced in \cite{254465} for efficient encoding and encryption, but it is totally different from our ternary quantization methods.

\begin{table*}[]
\centering
\caption{Comparison of encryption based privacy preserving FL systems}
\label{comparison}
\begin{tabular}{cccccc}
\toprule
\multirow{2}{*}{Proposed systems}     & \multicolumn{2}{c}{Threat model}    & \multicolumn{1}{l}{\multirow{2}{*}{Encryption scheme}} & \multicolumn{1}{l}{\multirow{2}{*}{Without TTP}} & \multicolumn{1}{l}{\multirow{2}{*}{Threshold Based}} \\ \cline{2-3}
                                        & Server             & Client         & \multicolumn{1}{l}{}                                   & \multicolumn{1}{l}{}                                        & \multicolumn{1}{l}{}                              \\ \toprule
 Phong \emph{et al.}\cite{8241854}           & honest but curious & honest          & Paillier, LWE                                           & \XSolidBrush                                    & \XSolidBrush                        \\ \hline
Truex \emph{et al.}\cite{10.1145/3338501.3357370}                            & honest but curious & majority honest & Paillier                                                &\XSolidBrush                                    & \CheckmarkBold                         \\ \hline
Xu \emph{et al.}\cite{10.1145/3338501.3357371}                       & honest but curious & majority honest & functional encryption                                                &\XSolidBrush                                    & \XSolidBrush                         \\ \hline
Batchcrypt\cite{254465}                              & honest but curious &  honest & Paillier                                                & \CheckmarkBold                                    & \XSolidBrush                      \\ \hline
DAEQ-FL(our system) & honest but curious & majority honest & ElGamal                                                 & \CheckmarkBold                                   & \CheckmarkBold                         \\ \bottomrule
\end{tabular}
\end{table*}

\section{Experimental Results}\label{Experiment Results}
In this section, we first introduce all experimental settings, followed by the encryption cost and time consumption. Finally, the results of model performances will be discussed.

\subsection{Simulation Settings}
In our simulations, we use CNN for MNIST \cite{lecun-mnisthandwrittendigit-2010} digit number classifications, ResNet for CIFAR10 \cite{alex} image classifications and stacked LSTM \cite{hochreiter1997long} for Shakespeare \cite{shakespeare} next word prediction task. All three datasets are non-iid among different clients.

MNIST is a 28x28 grey scale digit number image dataset containing 60,000 training images and 10,000 testing images with 10 different kinds of label classes (0$\sim$9). All the clients' training data are distributed according to their label classes and most clients contain only two kinds of digits for non-iid partition.

CIFAR10 contains 10 different kinds of 50,000 training and 10,000 testing 32x32x3 images. Similar to MNIST, the whole training data are horizontally sampled and each client owns five different kinds of object images.

The Shakespeare dataset is built from the whole work of Wailliams Shakespeare. It has in total 4,226,073 samples with 1129 role players and the data samples of each role player represent the dataset on each client. Additionally, 90\% of the user's data are randomly divided as the training data and the rest are testing data. This dataset is naturally non-iid and unbalanced, with some clients having few lines and others a large number of lines. In order to reduce the training time, we follow the method used in \cite{DBLP:journals/corr/abs-1812-01097} to randomly select 5\% of the total users and remove those containing less than 64 samples.

Note that we do not apply any data augmentation techniques \cite{tanner1987calculation,van2001art} to boost the final global model performances to reduce local computational complexity, since the main purpose of this work is not to achieve the state of the art model performances in FL; instead, we aim to present a distributed encryption method for better privacy preservation in FL without considerably increasing the computational and communication costs.

A CNN model is adopted to train on MNIST in the FL framework, which contains two 3x3 convolution layers with 32 and 64 filter channels, respectively, followed by a 2x2 max pooling layer. And then, a hidden layer with $128$ neurons is fully connected to the flattened output of the max pooling layer. Thus, the whole CNN model has 1,625,866 learnable parameters.

CIFAR10 dataset is to be learned by a ResNet model. The input images firstly pass through a 3x3 convolutional layer with 64 channels, followed by a batch normalization layer \cite{DBLP:journals/corr/IoffeS15} with the Relu \cite{DBLP:journals/corr/abs-1803-08375} activation function. Its output is connected to four sequentially connected block layers with 64, 128, 256, 512 filter channels, respectively. Each block layer contains two residual blocks containing two convolutional layers, each followed by a batch normalization layer and a shortcut connection. All the trainable parameters of the batch normalization layers are disabled, because they are observed to perform poorly with small batch sizes and non-iid data \cite{DBLP:journals/corr/IoffeS15,zhu2020federated}. The full ResNet model has 11,164,362 trainable parameters.

The Shakespeare dataset is trained by a stacked LSTM model which contains two LSTM layers, each with 256 neurons. Since we use the module cudnnLSTM of Tensorflow \cite{tensorflow2015-whitepaper} , the layer bias is twice as large as the original LSTM layer. Thus, the full model contains 819,920 parameters.

In the experiments training the CNN and ResNet for image classification, the FL system consists of a total of 20 clients, each containing 3000 and 2500 data samples for MNIST and CIFAR10, respectively. All the testing images are evenly and randomly distributed on all the clients. The total number of communication rounds is set to be 200 and all the clients are connected to the server in each round.
In the experiments for language modeling using the LSTM, we randomly sample 5\% of the entire role players (36 role users containing at least 64 samples) in Shakespeare dataset. In each round of FL learning, only 10 out of the 36 clients are randomly chosen to participate, following the settings in \cite{DBLP:journals/corr/abs-1812-01097}. The total number of communication rounds is set to be 100. Besides, the test accuracy of the global model is examined on local data only. Both correct predictions and test data size of connected clients are sent to the server to get the global test accuracy (note the server does not contain any test data).

The key size and group size of the distributed additive ElGamal encryption are set to 256 and 3072, respectively, to offer a 128-bit security level. Besides, the bit length for encoding is chosen to be from 2 to 15. The log recovery is used when the encoding bit length ranges from 2 to 10, whiles the brute force recovery method is adopted when the bit length is larger than 10.

We use the standard SGD algorithm for all model training. For the CNN models, the number of local epochs is set to 2, the batch size is 50, and the learning rate is 0.1 with a decay rate of 0.995 over the FL rounds. We do not use any momentum for training the CNNs, while the momentum is set to be 0.5 for the ResNet. For the LSTM, the local epoch is set to 1, the batch size is 10, and the learning rate is 0.5 with a decay rate of 0.995.

\subsection{Encryption Cost and Brute Force Recovery Time}
At first, we compare the communication costs between the proposed DAEQ-FL and a threshold based Paillier method in terms of encryption and recovery costs.
\begin{table*}[]
\centering
\caption{Communication costs of one connected client for both encryption and partial decryption with 128-bit security level, \# of ciphertexts means the number of transmitted ciphertexts for encryption and decryption, respectively.}
\label{enccc}
\begin{tabular}{ccccc}
\toprule
Models            & Enc Uploads (MB) & Dec Downloads (MB) & Dec Uploads (MB) & \# of Ciphertexts  \\
\toprule
CNN (DAEQ-FL)       & 0.3876+0.0059    & 0.0059             & 0.0029          &  16+24 \\
ResNet (DAEQ-FL)    & 2.6618+0.0161    & 0.0161             & 0.0081          & 44+66  \\
LSTM (DAEQ-FL)      & 0.1955+0.0066    & 0.0066             & 0.0033          & 18+27 \\
CNN (Paillier)    & 595.4099         & 595.4099           & 595.4099         & 1625866+3251732 \\
ResNet (Paillier) & 4088.5115        & 4088.5115          & 4088.5115        & 11164362+22328724 \\
LSTM (Paillier)   & 300.2637         & 300.2637           & 300.2637     & 819920+1639840   \\
\bottomrule
\end{tabular}
\end{table*}

We experiment with three different models (CNN, ResNet and LSTM) in our DAEQ-FL system and compare them with the Paillier-based variants. As shown in Table \ref{enccc}, the communication cost of our system is dramatically less than those Paillier-based systems. The best case has been achieved in the LSTM, where each client consumes only 0.212MB of communication costs in one round, whereas the Pillier system takes 900.7911MB, which is about 4249x of our system. For training the CNN and ResNet, the proposed DEAQ-FL costs 0.4023MB and 2.7021MB, respectively, which accounts for approximately 0.023\% and 0.022\% of the Paillier based variants.

Next, we compare the runtime of the ElGamal encryption used in our system and the conventional Paillier method under the same security level. The runtimes of the two encryption methods for encrypting and decrypting one number are listed in Table \ref{elgpatime}, where both the key size of Paillier and the group size of ElGamal are 3072.
\begin{table}[]
\centering
\caption{Runtime for encryption and decryption of one number using ElGamal and Paillier}
\label{elgpatime}
\begin{tabular}{ccc}
\toprule
Algorithm & Enc Time (s) & Dec Time (s) \\
\toprule
ElGamal   & 0.0029       & 0.0015       \\
Paillier  & 0.0501       & 0.0141  \\
\bottomrule
\end{tabular}
\end{table}

\begin{figure*}[!t]
\begin{minipage}[t]{1\linewidth}
\centering
\subfigure[CNN for MNIST dataset]{
\begin{minipage}[b]{0.3\textwidth}
\includegraphics[width=5cm]{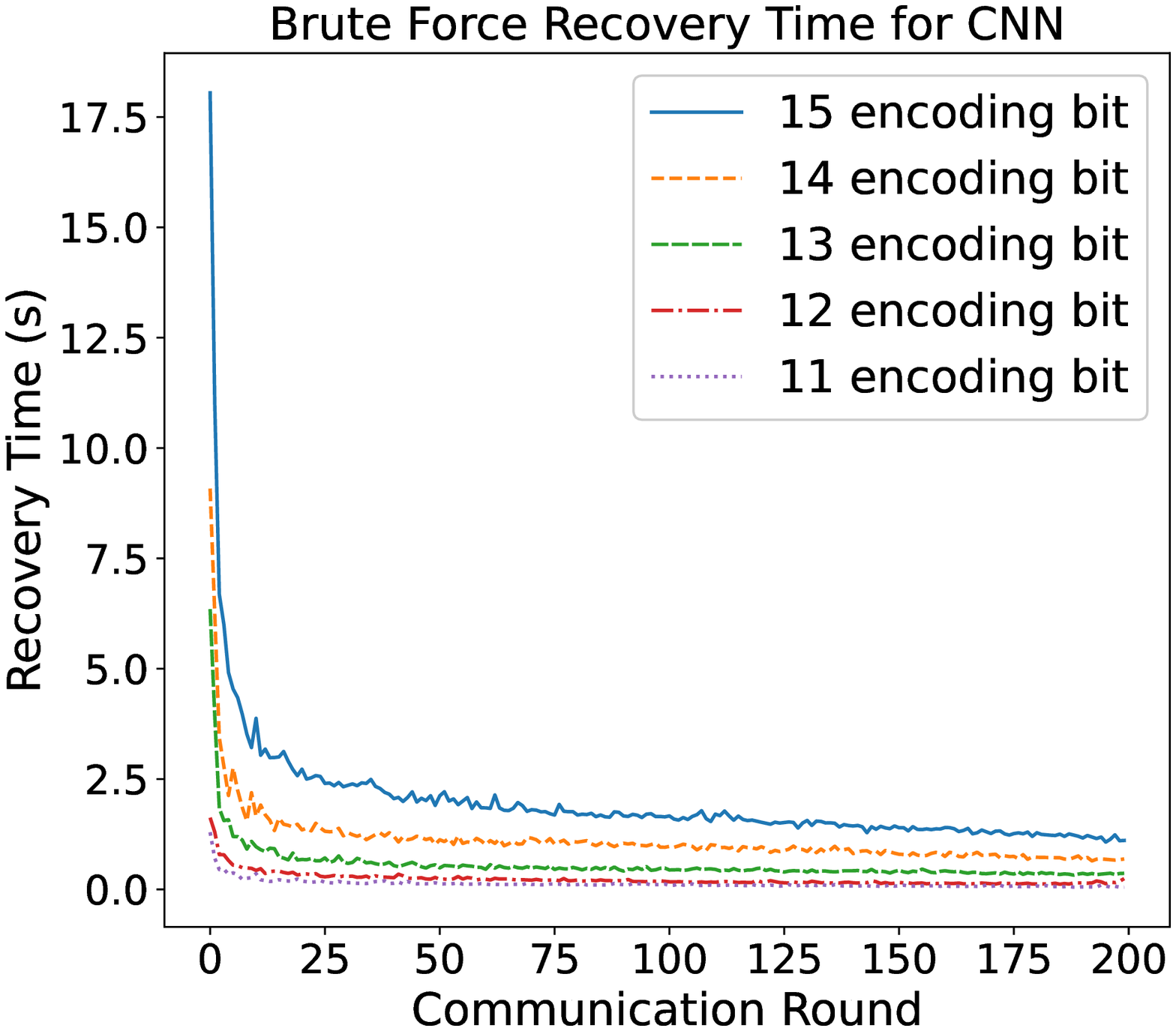}
\end{minipage}
}
\centering
\subfigure[ResNet for CIFAR10 dataset]{
\begin{minipage}[b]{0.3\textwidth}
\includegraphics[width=5cm]{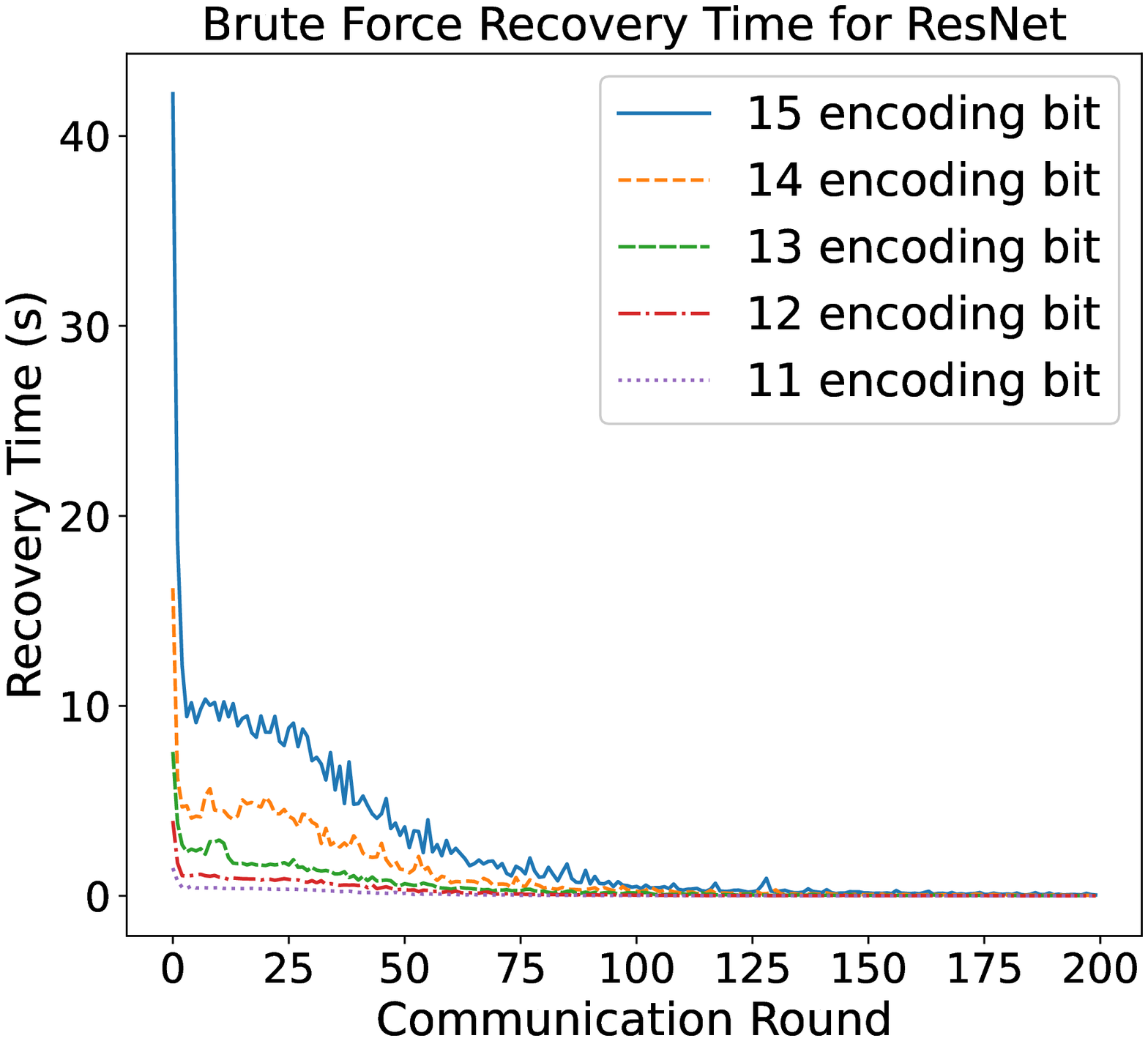}
\end{minipage}
}
\centering
\subfigure[LSTM for Shakespeare dataset]{
\begin{minipage}[b]{0.3\textwidth}
\includegraphics[width=5cm]{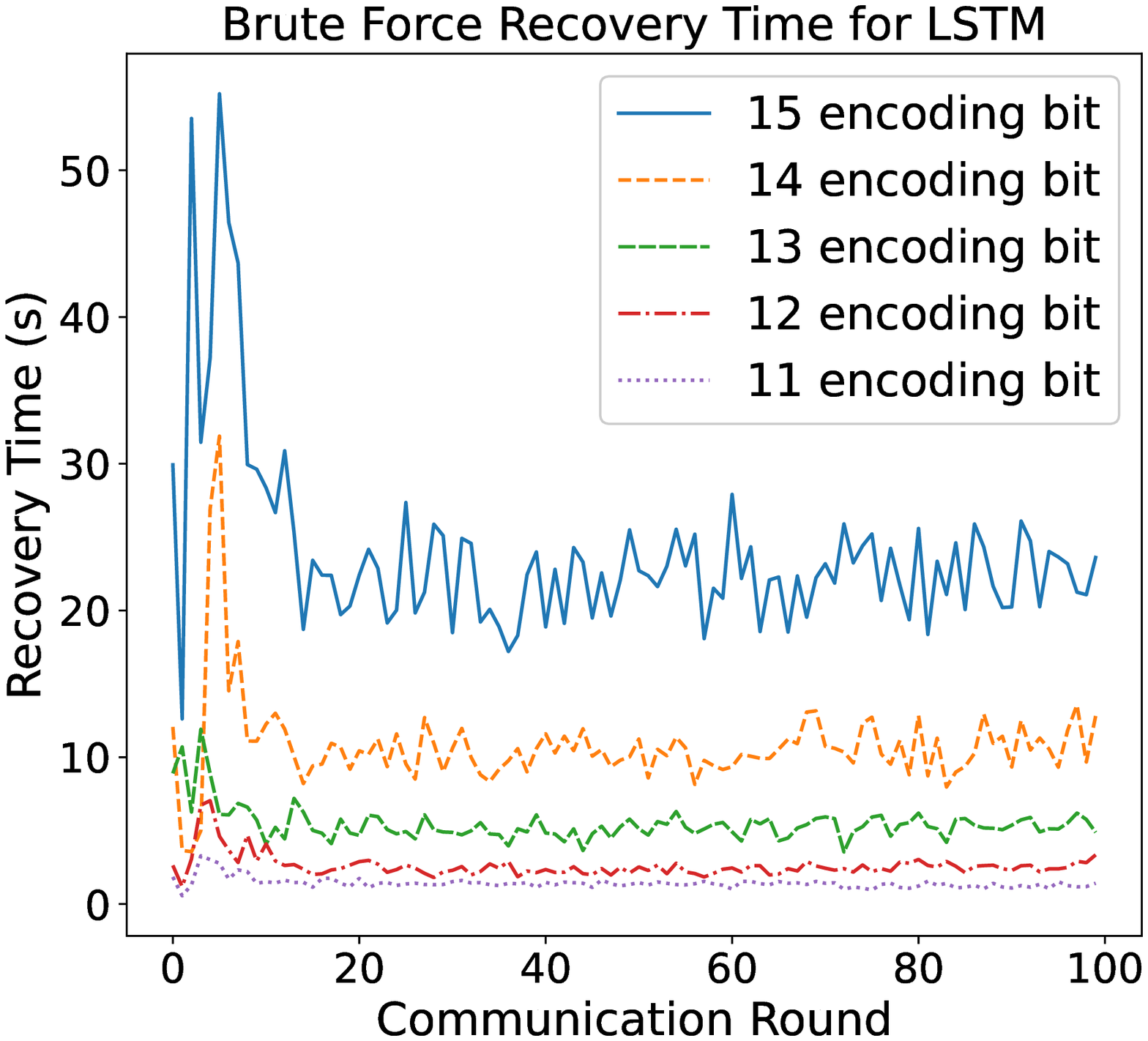}
\end{minipage}
}
\caption{Brute force recovery time with different encoding bit lengths for different learning models.}
\label{bftime}
\end{minipage}
\end{figure*}

From the results in Table \ref{elgpatime}, we can observe that ElGamal is approximately 17 times and 10 times faster than Paillier for encryption and decryption, respectively. However, since the Cramer transformation needs extra brute force recovery time, in the following, we explore the brute force recovery time corresponding to different encoding bit lengths for CNN, ResNet18 and stacked LSTM, respectively. The comparative results are plotted in Fig. \ref{bftime}.




Because the larger the encoding bit length is, the more time the brute force recovery will consume. Here, we experiment with the encoding bit length starting from 15 bits to 2 bits. The results clearly show that the difference in computation times goes bigger with rising of encoding bit. Specifically, CNN, ResNet and LSTM spend at most 18.0467s, 42.2203s and 55.2088s on recovery, respectively.


The CNN and ResNet show similar recovery time profile over the communication rounds. Their brute force recovery time are very large in the beginning, and quickly drop over the communication rounds. This is attributed to the fact that the gradients of the model parameters of the SGD decrease quickly as the global model parameters converge. It is surprisingly to see that the recovery time for the ResNet becomes almost zero, which is smaller than that of the CNN after approx. 100 communication rounds. This means the model gradients of the ResNet become very small at the end of federated model training.

By contrast, the recovery time of the LSTM does not drop to zero and keeps fluctuating at a relatively high level, especially for the 15-bit length encoding. There are two reasons for the above observations. First, training of the LSTM involves large gradients of recurrent connections, and those values are determined by the length of sequence. Second, the setting of the FL environment for the LSTM is very different from that of the CNN and ResNet, where only ten clients randomly participate global model aggregation in one communication round.

\begin{figure}[!t]
\begin{minipage}[t]{1\linewidth}
\centering
\subfigure[15 bit Brute Force Recovery]{
\begin{minipage}[b]{0.42\textwidth}
\includegraphics[width=7cm]{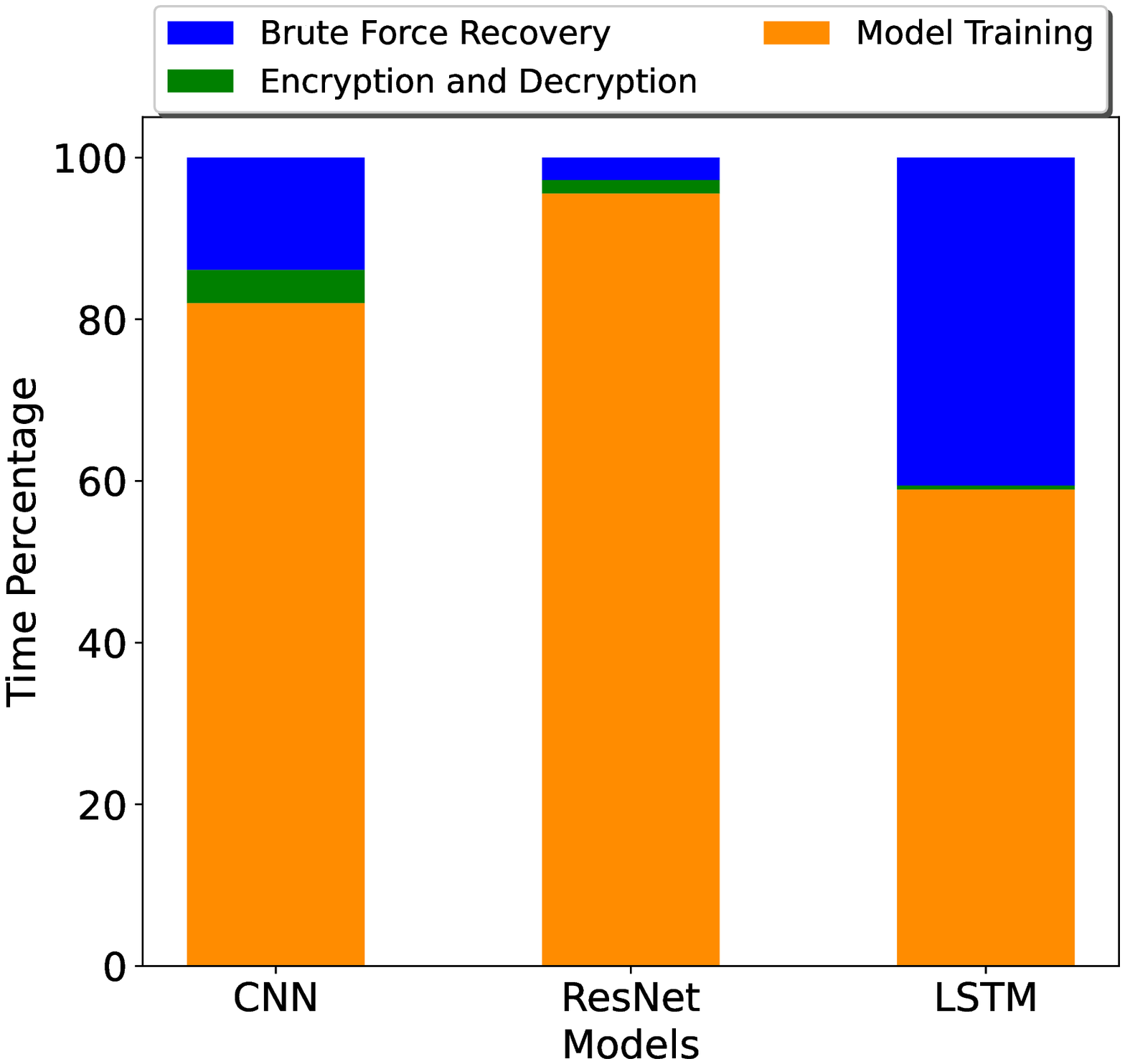}
\end{minipage}
}
\centering
\subfigure[10 bit Log Recovery]{
\begin{minipage}[b]{0.42\textwidth}
\includegraphics[width=7cm]{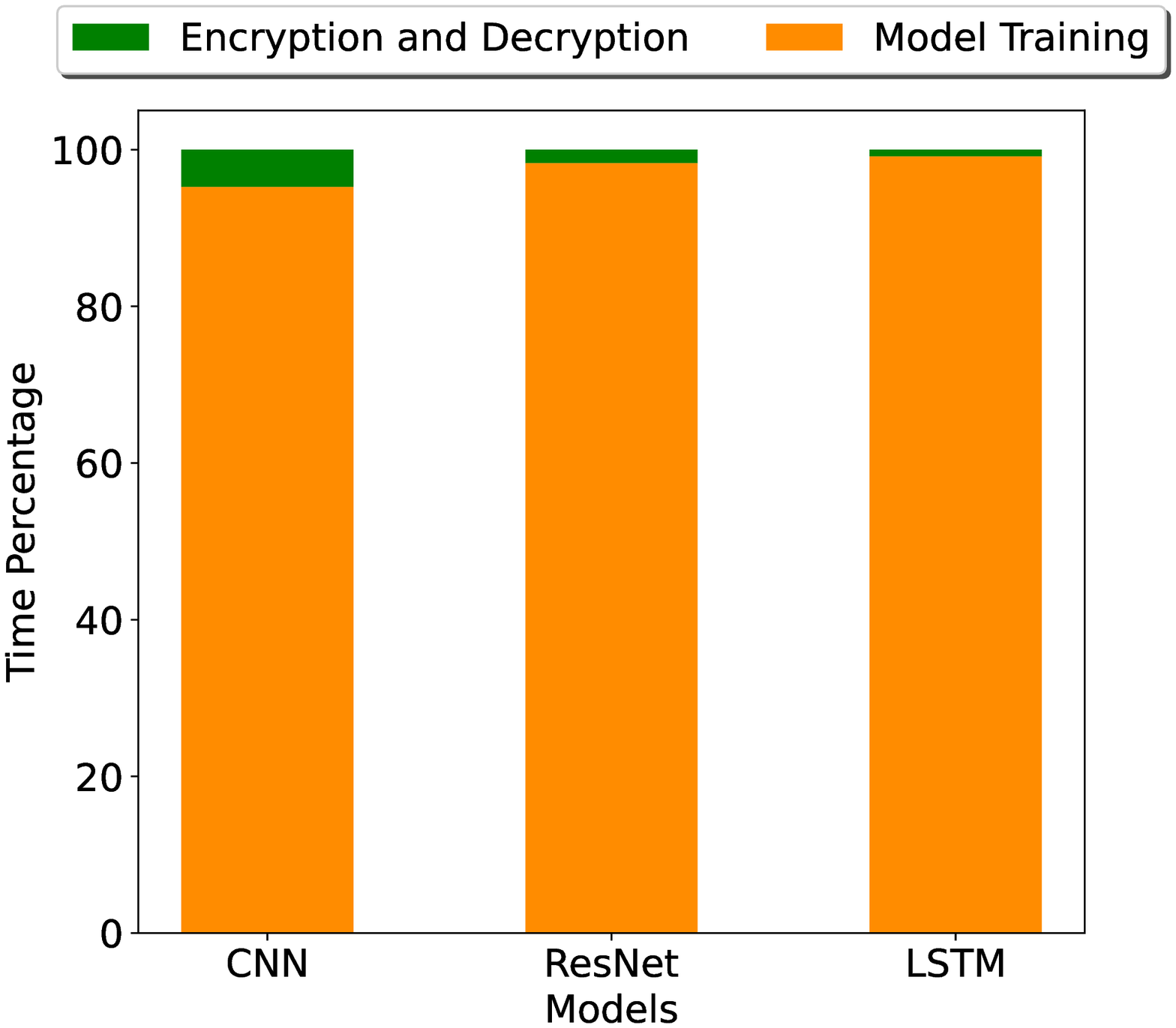}
\end{minipage}
}
\caption{Ratio of consumed time for model training, encryption and decryption, and brute force recovery for the 15-bit encoding length (a), and for the log recovery for the 10-bit encoding length (b) in one communication round.}
\label{percentage}
\end{minipage}
\end{figure}

Table \ref{bftime1} presents the time consumption of the brute force recovery for 15-bit encoding length. From Fig. \ref{percentage}(a) we can find that the average brute force recovery time accounts for a great proportion of the total elapsed time in each communication round, especially for the LSTM.
\begin{table}[]
\centering
\caption{Brute force recovery time for 15 encoding bit length}
\label{bftime1}
\begin{tabular}{cccc}
\toprule
Models & Max (s) & Min (s) & Avg (s) \\
\toprule
CNN    & 18.0467 & 1.0717  & 1.9786  \\
ResNet & 42.2203 & 0.0474  & 2.6491  \\
LSTM   & 55.2088 & 12.6166 & 23.8876 \\
\bottomrule
\end{tabular}
\end{table}

Therefore, we can use the log recovery instead of the brute force recovery when the encoding bit length is smaller than or equal to 10 so that the recovery time and the encryption time become negligible (Fig. \ref{percentage}(b)). Since the group size of $p$ is 3072 bit and $g_0$ is 2, the log recovery is not recommended when the plaintext message is larger than 3072 bit (11-bit encoding length). In order to avoid overflow, the maximum encoding bit length is set to be 10 in our simulations and the global performance drop caused by a low encoding bit length will be discussed in the next section.

\subsection{Learning Performance}
\begin{figure}[!t]
\begin{minipage}[t]{1\linewidth}
\centering
\subfigure[CNN for MNIST dataset]{
\begin{minipage}[b]{0.46\textwidth}
\includegraphics[width=1\textwidth]{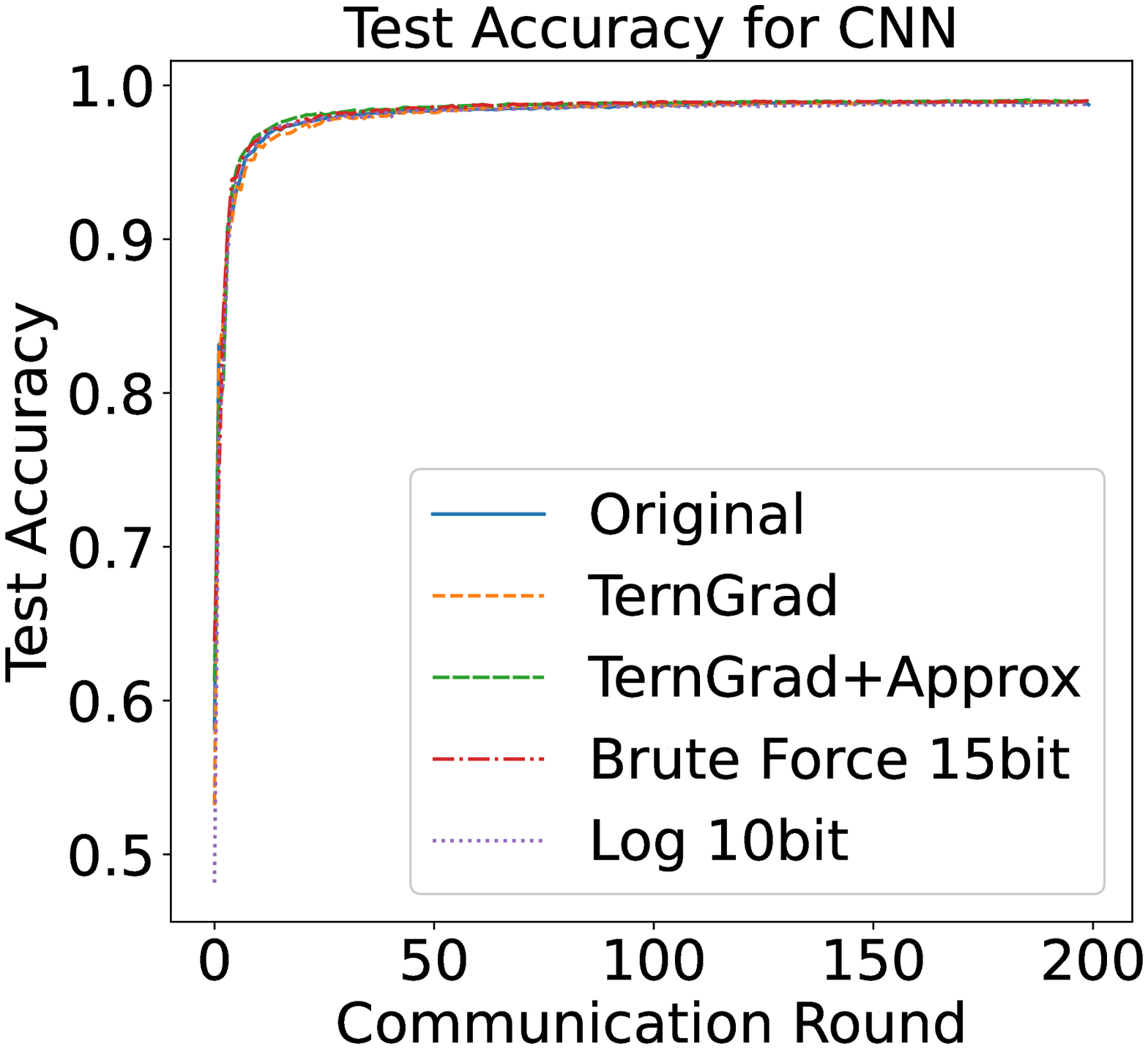}
\end{minipage}
}
\centering
\subfigure[ResNet for CIFAR10 dataset]{
\begin{minipage}[b]{0.46\textwidth}
\includegraphics[width=1\textwidth]{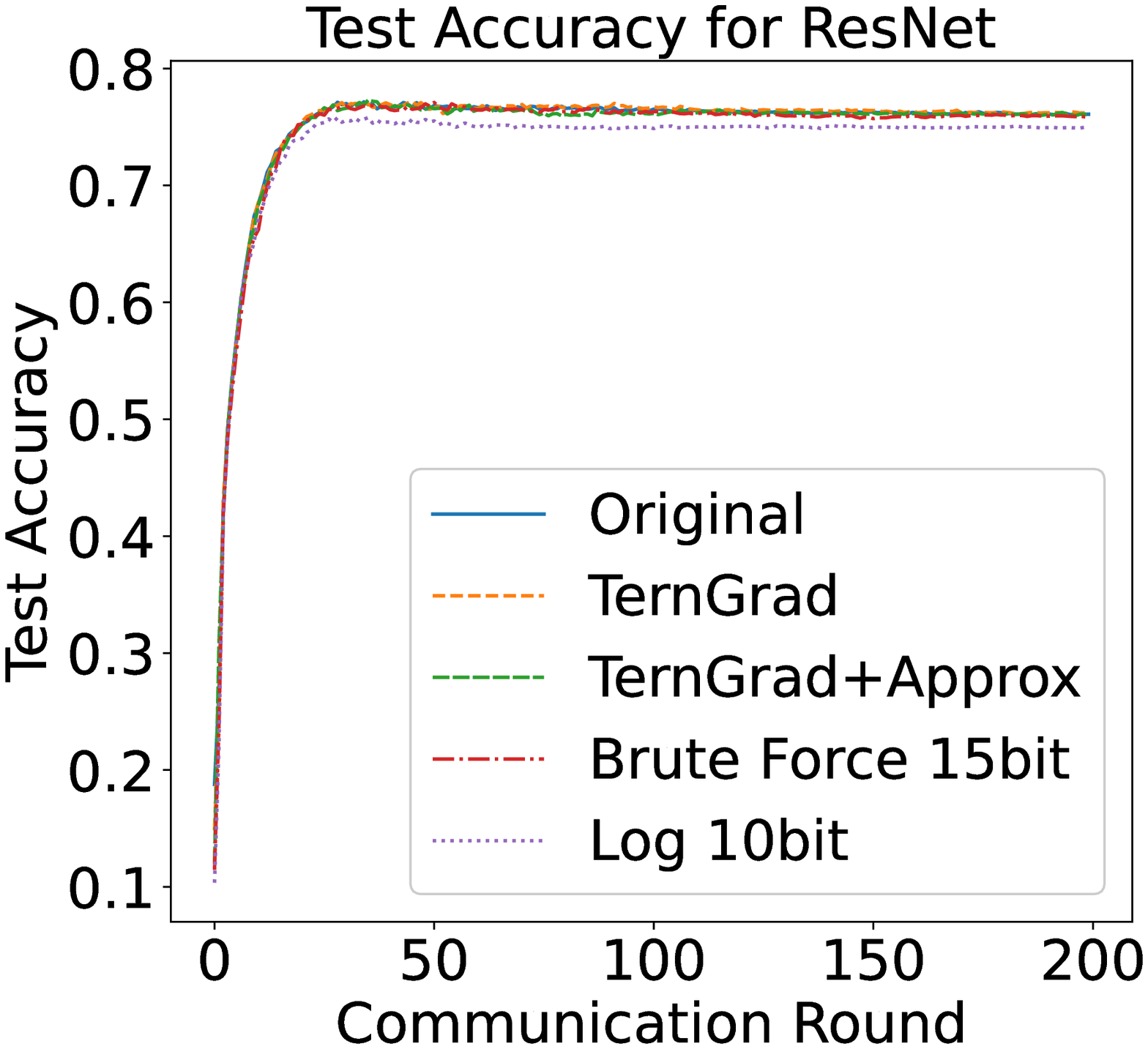}
\end{minipage}
}
\centering
\subfigure[LSTM for Shakespeare dataset]{
\begin{minipage}[b]{0.46\textwidth}
\includegraphics[width=1\textwidth]{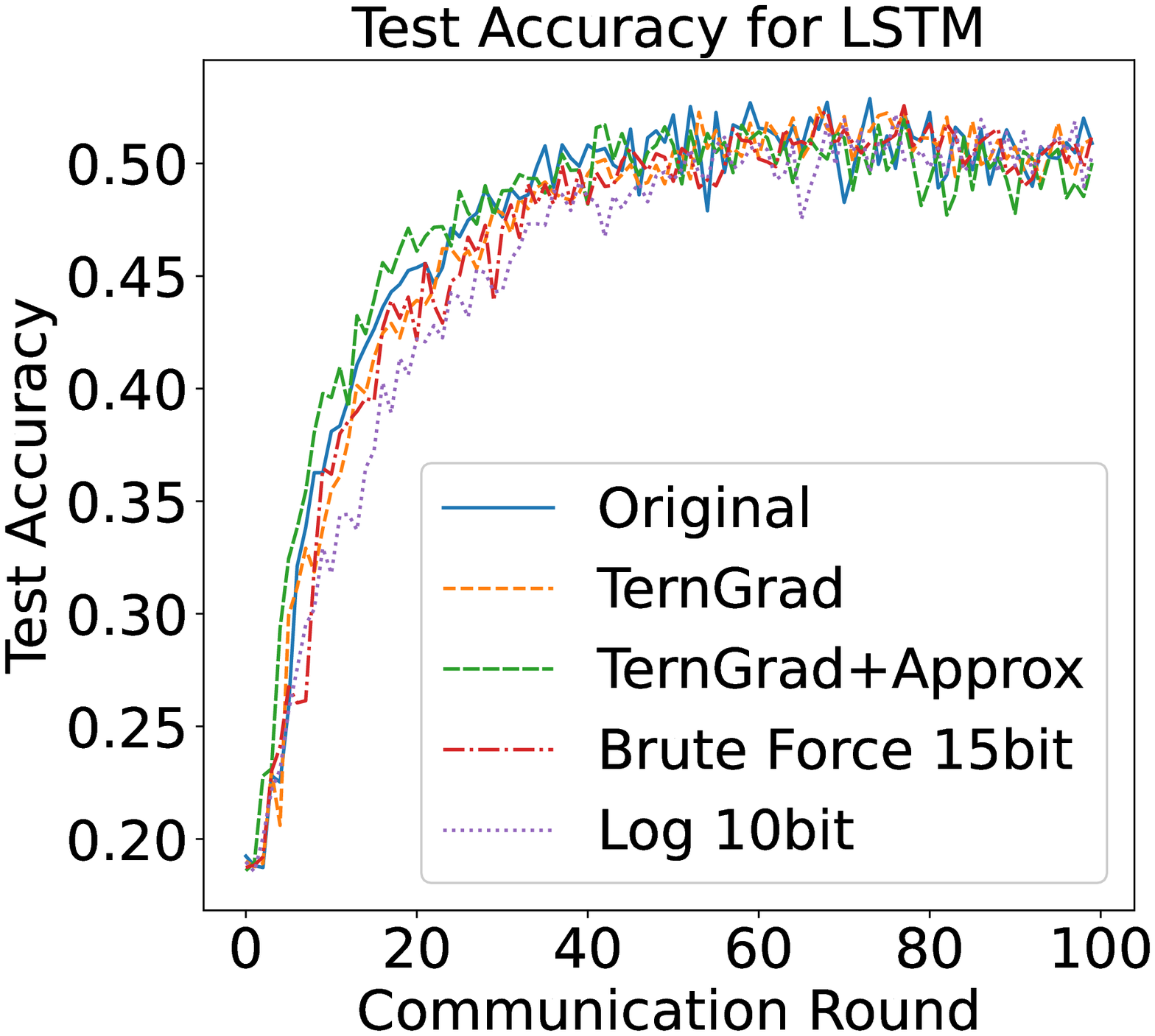}
\end{minipage}
}
\caption{The test accuracy of the global model for CNN, ResNet and LSTM in five different settings.}
\label{noenctestacc}
\end{minipage}
\end{figure}

In this section, we empirically examine influence of the TernGrad quantization, approximate aggregation and encoding length on the learning performance of the proposed DAEQ-FL system. Fig. \ref{noenctestacc} shows the test accuracy of the three models with or without encryption operations. For non encryption cases, `Original' represents standard FL, `TernGrad' means only quantization is used and `TernGrad+Approx' uses both quantization and approximated aggregation technique. And for encryption cases, brute force recovery is used for 15 encoding bit length cases and log recovery is adopted for 10 bit length scenarios.

From these results, we can see that the test accuracy of the models of the original FL and four variants of the DEAG-FL have achieved almost the same performance (in particular the CNN, with 98.97\% test accuracy). These results indicate that both the quantization and approximated aggregation have negligible impact on the test performance of the global model. Besides, neither the quantization or encryption has considerably slowed down the convergence speed over the communication rounds. It also can be seen that, the CNN and ResNet converge around the 25th round, while the LSTM is convergent around the 50th round.

More specifically, when the CNN trained on the MNIST dataset, the model performance does not degrade when the encoding bit length is decreased to 10. However, For the ResNet trained on CIFAR10, the test accuracy of the global model is reduce to 74.96\% by using 10-bit encoding, which is approximately 1\% lower than models without using encryption. The test accuracy of the LSTM models fluctuates between around 48\% and 52\%, regardless whether quantization or encryption are used or not. This can be due to the same reasons as discussed above.

\begin{figure}[!t]
\begin{minipage}[t]{1\linewidth}
\centering
\subfigure[CNN for MNIST dataset]{
\begin{minipage}[b]{0.46\textwidth}
\includegraphics[width=1\textwidth]{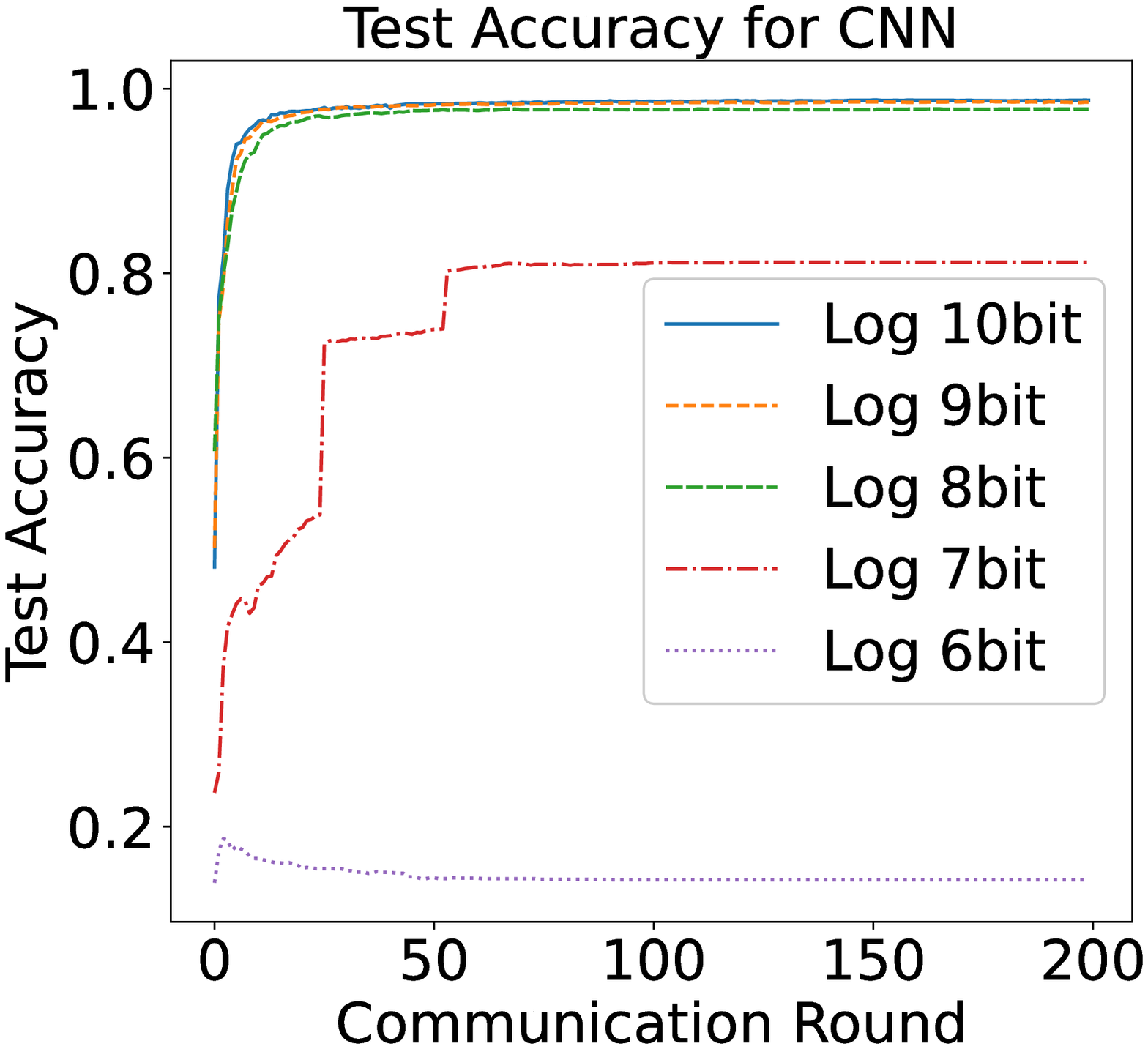}
\end{minipage}
}
\centering
\subfigure[ResNet for CIFAR10 dataset]{
\begin{minipage}[b]{0.46\textwidth}
\includegraphics[width=1\textwidth]{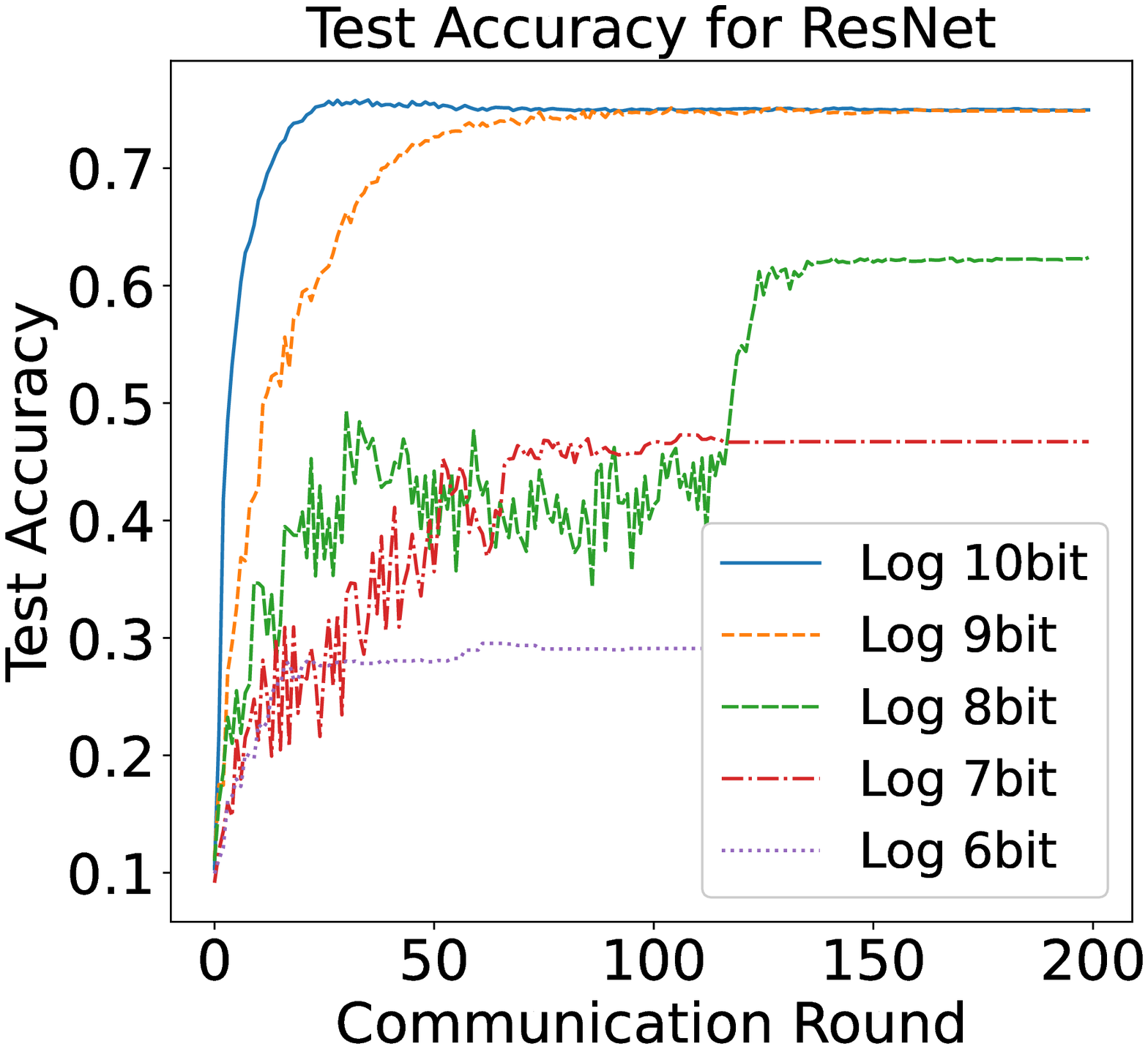}
\end{minipage}
}
\centering
\subfigure[LSTM for Shakespeare dataset]{
\begin{minipage}[b]{0.46\textwidth}
\includegraphics[width=1\textwidth]{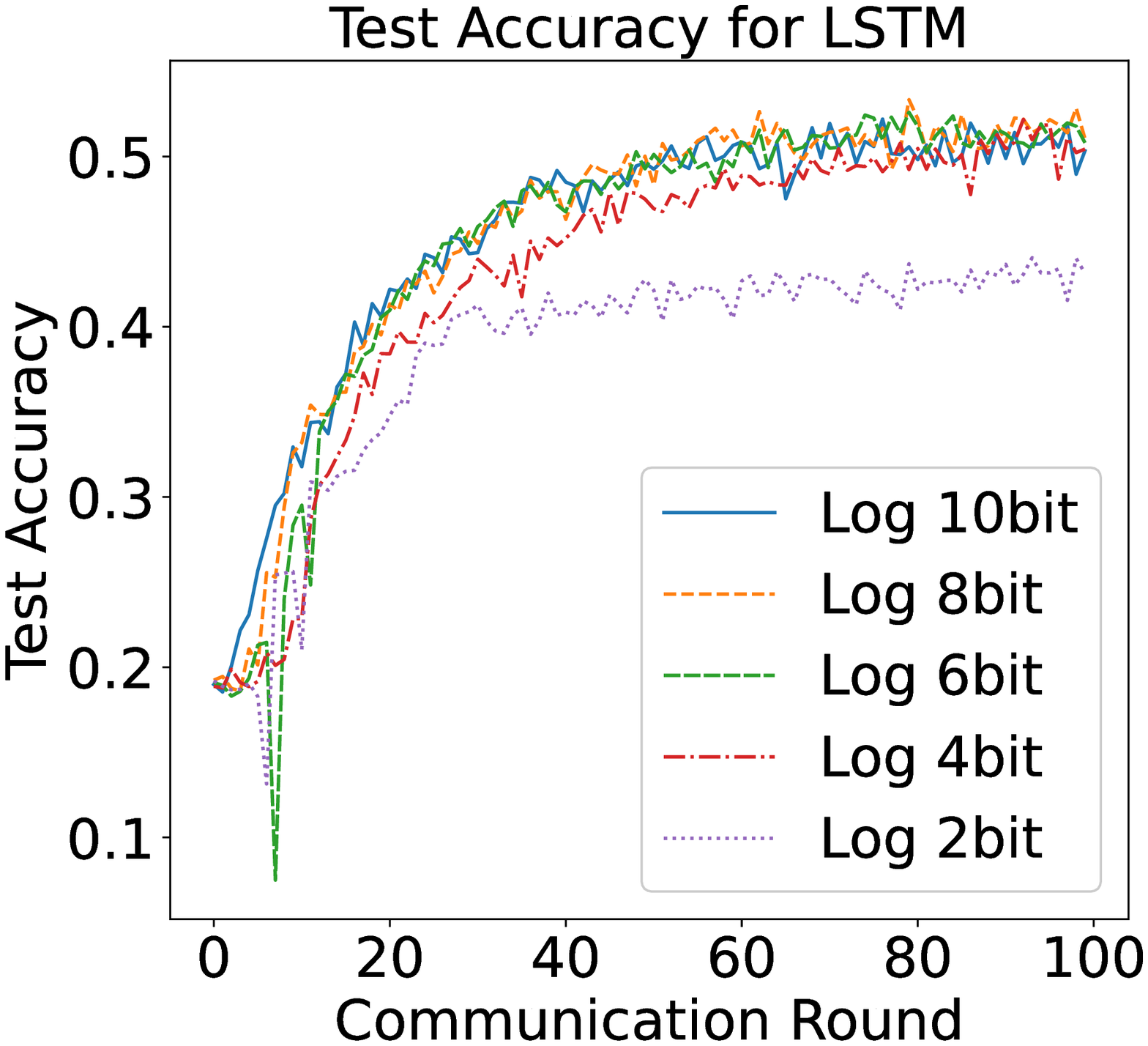}
\end{minipage}
}
\caption{The test accuracy of the global model with different encoding lengths.}
\label{logtestacc}
\end{minipage}
\end{figure}

Now we take a closer look at the learning performance of the global model of the proposed DAEQ-FL when the encoding bit length further decreases. The results for all models are shown in Fig. \ref{logtestacc}. It can be noticed that the convergence speed of the CNN and ResNet starts to decrease when the encoding bit length is smaller than 9. Besides, the model performance reduces dramatically when the encoding length is reduced to 7 bits or lower for CNN and to 8 bits or lower for the ResNet, respectively.

Reduction of the encoding bit length does not cause clear model degradation for the LSTM until when only 2 bits are used for encoding and the accuracy is reduced to 43.15\%. Nevertheless, we can still observe a slight drop in the convergence speed in the beginning of the communication rounds. The possible reason for this is that the model gradients of LSTM are large.

Overall, both TernGrad quantization and the approximate global model aggregation have little impact on the model performance, so long as the encoding bit length is not less than eight. Therefore, the proposed DEAQ-FL using the log recovery and an encoding length of 9 or 10 can achieve 128 bit security level with negligible degradation in performance and little increase in computational and communication costs for the CNN, ResNet and LSTM on the corresponding datasets studied in this work.

\section{Conclusion and Future Work}
In this paper, we propose a privacy-preserving solution that makes use of distributed key generation and additive ElGamal encryption to protect gradients in the federated learning framework. To reduce computational and communication costs, we also introduce ternary quantization of the local models and approximate aggregation of the global model, making our solution practical in complex machine learning models, such as deep neural networks, in the context of gradient encryption. The proposed DAEQ-FL system does not rely on a TTP for key pair generations, which enables the system to tolerate a certain number of malicious clients.

DEAQ-FL can adopt the computationally efficient log recovery when the encoding bit length is less than eleven, and there is no noticeable learning performance degradation when ten bits are used for encoding  (note only about 1\% accuracy loss for ResNet compared to the results without doing any encryption on the Shakespeare dataset). However, the model learning performance starts to clearly deteriorate when the encoding length is less than eight. Thus, brute force recovery must be adopted when the encoding length is larger than eleven. Although a large encoding length can enhance the coding precision, the amount of recovery time will considerably increase. According to our experimental results, the global models trained with a maximum of fifteen bits can perform comparably to the non-encrypted FL, while the recovery time is still acceptable. Since the model gradients tend to decrease to zero during training, the brute force recovery time is expected to become much smaller as the global model converges.

Although the proposed method shows highly promising performance for encrypted federated deep learning, it may need to balance a tradeoff between model performance and computation time needed for plaintext recovery when a small model (e.g., a logistic regression) is adopted. This is mainly because a smaller model will require a much longer encoding length (e.g., more than fifteen), making the brute force based recovery intractable. Therefore, our future work will be dedicated to the development of a distributed additive homomorphic encryption without recovery that can be used in federated learning systems.


%

\appendix
\section{Security definitions}\label{sedef}

\textbf{Definition 1} (Discrete Logarithm Hard Problem - DLHP) Discrete Logarithm is considered to be hard if
\[
Pr[DLog_{\mathcal{G}, \mathcal{A}}(\alpha)=1] \leq negl(\alpha)
\]

\textbf{Definition 2} (Computational Diffle-Hellman - CDH) CDH is considered to be hard if

\begin{equation}\nonumber
\begin{aligned}
Pr[\mathcal{A}(g,g^a,g^b)=(g^{ab})]\leq negl(\alpha)
 \end{aligned}
\end{equation}

\textbf{Definition 3} (Decisional Diffie-Hellman - DDH) DDH is considered to be hard if

\begin{equation}\nonumber
\begin{aligned}
| Pr[\mathcal{A}(g,g^a,g^b,g^c)=1] - Pr[\mathcal{A}(g,g^a,g^b&,g^{ab})=1] |\\
&\leq negl(\alpha)\\
 \end{aligned}
\end{equation}

\textbf{Definition 4} Indistinguishability - Chosen plaintext attacks (IND - CPA)

Consider a following game between an adversary $\mathcal{A}$ and a challenger:

\textbf{Set up:} Challenger generates public parameters $<g,p,q>$, public key $h$ and secret key $s$, then sends public parameters and $pk$ to the $\mathcal{A}$.

\textbf{Challenge:} $\mathcal{A}$ chooses two plaintexts $m_0$, $m_1$ with same length, then sends them to the challenger. Challenger randomly selects $b\in \{{0,1}\}$, encrypts $m_b$
\[
C = Enc(pk, m_b)
\]
and sends challenge ciphertext $C$ to $\mathcal{A}$. Finally, $\mathcal{A}$ would compute $b^{'}$.

A scheme is security against CPA if the advantage
\[
\textbf{Adv}^{CPA}_\mathcal{A}(\alpha)= \left|Pr(b=b^{'})-\frac{1}{2} \right|
\]
is negligible.

\section{Security analysis}\label{seana}

\begin{itemize}
\item Parameters generation: server runs a polynomial-time $\mathcal{G}(1^{\alpha})$ to generate public parameters $<\mathbb{G},g,p,q>$, where $g$ is a generator of $\mathbb{G}$ which is a cyclic group with prime order $q$, $p$ is a large prime number satisfying $q|p-1$, $y$ is a random element in $\mathbb{G}$ and $\alpha$ corresponding to security level is bit length of $q$. Given a polynomial-time algorithm $\mathcal{A}$ and a negligible function $negl$, the security of key generation and encryption are based on following definitions.

\item Key generation: $x\xleftarrow{\$} \mathbb{Z}_{q}$, $h=g^{x}\pmod{p}$.

\item Encryption: $g^{m}\in \mathbb{Z}_{p}^{*}$,  $r\xleftarrow{\$}\mathbb{Z}_{q}^{*}$, $<c_{1}=g^{r}\pmod{p}, c_{2}=g^{m}\cdot h^{r}\pmod{p}>$.

\item Decryption:
$\frac{c_{2}}{c_{1}^{x}}=\frac{mh^{r}}{(g^{r})^{x}}=\frac{g^{m}\cdot g^{xr}}{g^{rx}}=g^{m}\pmod{p}$

\end{itemize}

Additive ElGamal is CPA-secure against chosen ciphertext attacks if the advantage of $\mathcal{A}$ is negligible in $\alpha$ satisfying as follows:
\[
\textbf{Adv}^{CPA}_\mathcal{A}(\alpha)= \left|Pr(b=b^{'})-\frac{1}{2} \right|
\]
\begin{proof}

Ciphertexts under additive ElGamal:
\[
C = (c_{1}, c_{2}) = (g^{r}, g^{m_b}\cdot h^{r}) = g^{m_{b}+x\cdot r}
\]

$\mathcal{A}$ computes:
\[
C^{'} = C\cdot g^{m_{b^{'}}^{-1}}=g^{x\cdot r} \,\,\, iff \,\,\, b = b^{'}
\]
Besides, $\mathcal{A}$ wants to determine whether $C^{'}$ is DH agreement. $\mathcal{A}$ could "win" this game with a probability:

\begin{equation}\nonumber
\begin{aligned}
\textbf{Adv}^{ElG}_\mathcal{A}(\alpha)=||&Pr[\mathcal{A}(g,g^x,g^r,g^{x\cdot r+m_{b}-m_{b^{'}}})=1] \\
&- Pr[\mathcal{A}(g,g^x,g^r,g^{x\cdot r})=1]|-\frac{1}{2} |\\
&\leq negl(\alpha)\\
\end{aligned}
\end{equation}

\textbf{Theorem 1} The exponential ElGamal and distribute key generation are CPA-secure and  correctness-satisfying according to \cite{cramer1997secure} and \cite{gennaro1999secure}.
\end{proof}

\bibliographystyle{unsrt}
\bibliography{references}

\begin{thebibliography}{10}

\bibitem{mcmahan2017communication}
Brendan McMahan, Eider Moore, Daniel Ramage, Seth Hampson, and Blaise~Aguera
  y~Arcas.
\newblock Communication-efficient learning of deep networks from decentralized
  data.
\newblock In {\em Artificial Intelligence and Statistics}, pages 1273--1282.
  PMLR, 2017.

\bibitem{LI2020106854}
Li~Li, Yuxi Fan, Mike Tse, and Kuo-Yi Lin.
\newblock A review of applications in federated learning.
\newblock {\em Computers \& Industrial Engineering}, 149:106854, 2020.

\bibitem{yang2019federated}
Qiang Yang, Yang Liu, Tianjian Chen, and Yongxin Tong.
\newblock Federated machine learning: Concept and applications.
\newblock {\em ACM Transactions on Intelligent Systems and Technology (TIST)},
  10(2):1--19, 2019.

\bibitem{krizhevsky2017imagenet}
Alex Krizhevsky, Ilya Sutskever, and Geoffrey~E Hinton.
\newblock Imagenet classification with deep convolutional neural networks.
\newblock {\em Communications of the ACM}, 60(6):84--90, 2017.

\bibitem{8945292}
Y.~{Chen}, X.~{Sun}, and Y.~{Jin}.
\newblock Communication-efficient federated deep learning with layerwise
  asynchronous model update and temporally weighted aggregation.
\newblock {\em IEEE Transactions on Neural Networks and Learning Systems},
  31(10):4229--4238, 2020.

\bibitem{zhu2020real}
Hangyu Zhu and Yaochu Jin.
\newblock Real-time federated evolutionary neural architecture search.
\newblock {\em arXiv preprint arXiv:2003.02793}, 2020.

\bibitem{ribero2020}
Monica Ribero and Haris Vikalo.
\newblock Communication-efficient federated learning via optimal client
  sampling.
\newblock {\em arXiv:2007.15197v2}, 2020.

\bibitem{amiri2020federated}
Mohammad~Mohammadi Amiri, Deniz Gunduz, Sanjeev~R Kulkarni, and H~Vincent Poor.
\newblock Federated learning with quantized global model updates.
\newblock {\em arXiv preprint arXiv:2006.10672}, 2020.

\bibitem{xu2020ternary}
Jinjin Xu, Wenli Du, Yaochu Jin, Wangli He, and Ran Cheng.
\newblock Ternary compression for communication-efficient federated learning.
\newblock {\em IEEE Transactions on Neural Networks and Learning Systems},
  2020.

\bibitem{zhu2016trained}
Chenzhuo Zhu, Song Han, Huizi Mao, and William~J Dally.
\newblock Trained ternary quantization.
\newblock {\em arXiv preprint arXiv:1612.01064}, 2016.

\bibitem{dai2019hyper}
Xinyan Dai, Xiao Yan, Kaiwen Zhou, Kelvin~KW Ng, James Cheng, and Yu~Fan.
\newblock Hyper-sphere quantization: Communication-efficient sgd for federated
  learning.
\newblock {\em arXiv preprint arXiv:1911.04655}, 2019.

\bibitem{du2020high}
Yuqing Du, Sheng Yang, and Kaibin Huang.
\newblock High-dimensional stochastic gradient quantization for
  communication-efficient edge learning.
\newblock {\em IEEE Transactions on Signal Processing}, 68:2128--2142, 2020.

\bibitem{shokri2015privacy}
Reza Shokri and Vitaly Shmatikov.
\newblock Privacy-preserving deep learning.
\newblock In {\em the 22nd ACM CCS}.

\bibitem{orekondy2018gradient}
Tribhuvanesh Orekondy, Seong~Joon Oh, Yang Zhang, Bernt Schiele, and Mario
  Fritz.
\newblock Gradient-leaks: Understanding and controlling deanonymization in
  federated learning.
\newblock {\em arXiv preprint arXiv:1805.05838}, 2018.

\bibitem{8241854}
L.~T. {Phong}, Y.~{Aono}, T.~{Hayashi}, L.~{Wang}, and S.~{Moriai}.
\newblock Privacy-preserving deep learning via additively homomorphic
  encryption.
\newblock {\em IEEE Transactions on Information Forensics and Security},
  13(5):1333--1345, 2018.

\bibitem{geiping2020inverting}
Jonas Geiping, Hartmut Bauermeister, Hannah Dröge, and Michael Moeller.
\newblock Inverting gradients -- how easy is it to break privacy in federated
  learning?, 2020.

\bibitem{8712695}
H.~{Li} and T.~{Han}.
\newblock An end-to-end encrypted neural network for gradient updates
  transmission in federated learning.
\newblock In {\em 2019 Data Compression Conference (DCC)}, pages 589--589,
  2019.

\bibitem{10.1007/978-3-540-79228-4_1}
Cynthia Dwork.
\newblock Differential privacy: A survey of results.
\newblock In Manindra Agrawal, Dingzhu Du, Zhenhua Duan, and Angsheng Li,
  editors, {\em Theory and Applications of Models of Computation}, pages 1--19,
  Berlin, Heidelberg, 2008. Springer Berlin Heidelberg.

\bibitem{gentry2009fully}
Craig Gentry and Dan Boneh.
\newblock {\em A fully homomorphic encryption scheme}, volume~20.
\newblock Stanford University Stanford, 2009.

\bibitem{abadi2016deep}
Martin Abadi, Andy Chu, Ian Goodfellow, H~Brendan McMahan, Ilya Mironov, Kunal
  Talwar, and Li~Zhang.
\newblock Deep learning with differential privacy.
\newblock In {\em The 2016 ACM CCS}.

\bibitem{geyer2017differentially}
Robin~C Geyer, Tassilo Klein, and Moin Nabi.
\newblock Differentially private federated learning: A client level
  perspective.
\newblock {\em arXiv preprint arXiv:1712.07557}, 2017.

\bibitem{wei2020federated}
Kang Wei, Jun Li, Ming Ding, Chuan Ma, Howard~H Yang, Farhad Farokhi, Shi Jin,
  Tony~QS Quek, and H~Vincent Poor.
\newblock Federated learning with differential privacy: Algorithms and
  performance analysis.
\newblock {\em IEEE Transactions on Information Forensics and Security}, 2020.

\bibitem{Paillier1999public}
Pascal Paillier.
\newblock Public-key cryptosystems based on composite degree residuosity
  classes.
\newblock In {\em TAMC}. Springer, 1999.

\bibitem{10.1145/3338501.3357370}
Stacey Truex, Nathalie Baracaldo, Ali Anwar, Thomas Steinke, Heiko Ludwig, Rui
  Zhang, and Yi~Zhou.
\newblock A hybrid approach to privacy-preserving federated learning.
\newblock In {\em the 12th ACM AISec}, New York, NY, USA. Association for
  Computing Machinery.

\bibitem{damgaard2001generalisation}
Ivan Damg{\aa}rd and Mads Jurik.
\newblock A generalisation, a simpli. cation and some applications of
  paillier's probabilistic public-key system.
\newblock In {\em PKC}. Springer, 2001.

\bibitem{mandal2019privfl}
Kalikinkar Mandal and Guang Gong.
\newblock Privfl: Practical privacy-preserving federated regressions on
  high-dimensional data over mobile networks.
\newblock In {\em The 2019 ACM CCSW}.

\bibitem{hao2019towards}
Meng Hao, Hongwei Li, Guowen Xu, Sen Liu, and Haomiao Yang.
\newblock Towards efficient and privacy-preserving federated deep learning.
\newblock In {\em ICC 2019}. IEEE.

\bibitem{254465}
Chengliang Zhang, Suyi Li, Junzhe Xia, Wei Wang, Feng Yan, and Yang Liu.
\newblock Batchcrypt: Efficient homomorphic encryption for cross-silo federated
  learning.
\newblock In {\em 2020 {USENIX} {ATC}}, July.

\bibitem{8825829}
L.~{Zhao}, Q.~{Wang}, Q.~{Zou}, Y.~{Zhang}, and Y.~{Chen}.
\newblock Privacy-preserving collaborative deep learning with unreliable
  participants.
\newblock {\em IEEE Transactions on Information Forensics and Security},
  15:1486--1500, 2020.

\bibitem{8747377}
M.~{Kim}, J.~{Lee}, L.~{Ohno-Machado}, and X.~{Jiang}.
\newblock Secure and differentially private logistic regression for
  horizontally distributed data.
\newblock {\em IEEE Transactions on Information Forensics and Security},
  15:695--710, 2020.

\bibitem{ElGamal1985public}
Taher ElGamal.
\newblock A public key cryptosystem and a signature scheme based on discrete
  logarithms.
\newblock {\em IEEE Transactions on Information Theory}, 31(4):469--472, 1985.

\bibitem{goodfellow2016deep}
Ian Goodfellow, Yoshua Bengio, Aaron Courville, and Yoshua Bengio.
\newblock {\em Deep learning}, volume~1.
\newblock MIT Press, Cambridge, 2016.

\bibitem{lecun2015deep}
Yann LeCun, Yoshua Bengio, and Geoffrey Hinton.
\newblock Deep learning.
\newblock {\em Nature}, 521(7553):436--444, 2015.

\bibitem{schmidhuber2015deep}
J{\"u}rgen Schmidhuber.
\newblock Deep learning in neural networks: An overview.
\newblock {\em Neural Networks}, 61:85--117, 2015.

\bibitem{deng2014deep}
Li~Deng and Dong Yu.
\newblock Deep learning: methods and applications.
\newblock {\em Foundations and Trends in Signal Processing}, 7(3--4):197--387,
  2014.

\bibitem{nielsen2015neural}
Michael~A Nielsen.
\newblock {\em Neural networks and deep learning}, volume 2018.
\newblock Determination press San Francisco, CA, 2015.

\bibitem{lecun1995convolutional}
Yann LeCun, Yoshua Bengio, et~al.
\newblock Convolutional networks for images, speech, and time series.
\newblock {\em The Handbook of Brain Theory and Neural Networks},
  3361(10):1995, 1995.

\bibitem{schuster1997bidirectional}
Mike Schuster and Kuldip~K Paliwal.
\newblock Bidirectional recurrent neural networks.
\newblock {\em IEEE Transactions on Signal Processing}, 45(11):2673--2681,
  1997.

\bibitem{moller1993scaled}
Martin~Fodslette M{\o}ller.
\newblock A scaled conjugate gradient algorithm for fast supervised learning.
\newblock {\em Neural Networks}, 6(4):525--533, 1993.

\bibitem{article}
Elaine Barker, William Barker, William Burr, William Polk, and Miles Smid.
\newblock Recommendation for key management – part 1: General (revision 3).
\newblock {\em NIST Special Publication Revision}, 3, 01 2005.

\bibitem{Knirsch20a}
Fabian Knirsch, Andreas Unterweger, Maximilian Unterrainer, and Dominik Engel.
\newblock {Comparison of the Paillier and ElGamal Cryptosystems for Smart Grid
  Aggregation Protocols}.
\newblock In {\em Proceedings of the 6th International Conference on
  Information Systems Security and Privacy (ICISSP)}, pages 232--239, Valetta,
  Malta, 2020. SciTePress.

\bibitem{cramer1997secure}
Ronald Cramer, Rosario Gennaro, and Berry Schoenmakers.
\newblock A secure and optimally efficient multi-authority election scheme.
\newblock {\em European Transactions on Telecommunications}, 8(5):481--490,
  1997.

\bibitem{shamir1979share}
Adi Shamir.
\newblock How to share a secret.
\newblock {\em Communications of the ACM}, 22(11):612--613, 1979.

\bibitem{feldman1987practical}
Paul Feldman.
\newblock A practical scheme for non-interactive verifiable secret sharing.
\newblock In {\em SFCS 1987}. IEEE.

\bibitem{pedersen1991non}
Torben~Pryds Pedersen.
\newblock Non-interactive and information-theoretic secure verifiable secret
  sharing.
\newblock In {\em CRYTO}. Springer, 1991.

\bibitem{gennaro1999secure}
Rosario Gennaro, Stanis{\l}aw Jarecki, Hugo Krawczyk, and Tal Rabin.
\newblock Secure distributed key generation for discrete-log based
  cryptosystems.
\newblock In {\em Eurocrypt}. Springer, 1999.

\bibitem{berrut2004barycentric}
Jean-Paul Berrut and Lloyd~N Trefethen.
\newblock Barycentric lagrange interpolation.
\newblock {\em SIAM Review}, 46(3):501--517, 2004.

\bibitem{wen2017terngrad}
Wei Wen, Cong Xu, Feng Yan, Chunpeng Wu, Yandan Wang, Yiran Chen, and Hai Li.
\newblock Terngrad: Ternary gradients to reduce communication in distributed
  deep learning.
\newblock In {\em NIPS}, 2017.

\bibitem{uspensky1937introduction}
James~Victor Uspensky.
\newblock Introduction to mathematical probability.
\newblock 1937.

\bibitem{truex2019hybrid}
Stacey Truex, Nathalie Baracaldo, Ali Anwar, Thomas Steinke, Heiko Ludwig, Rui
  Zhang, and Yi~Zhou.
\newblock A hybrid approach to privacy-preserving federated learning.
\newblock In {\em The 12th ACM Workshop on AISec}.

\bibitem{10.1145/3338501.3357371}
Runhua Xu, Nathalie Baracaldo, Yi~Zhou, Ali Anwar, and Heiko Ludwig.
\newblock Hybridalpha: An efficient approach for privacy-preserving federated
  learning.
\newblock In {\em The 12th ACM AISec}, New York, NY, USA, 2019. Association for
  Computing Machinery.

\bibitem{lecun-mnisthandwrittendigit-2010}
Yann LeCun and Corinna Cortes.
\newblock {MNIST} handwritten digit database.
\newblock 2010.

\bibitem{alex}
Alex Krizhevsky, Vinod Nair, and Geoffrey Hinton.
\newblock Cifar-10 (canadian institute for advanced research).

\bibitem{hochreiter1997long}
Sepp Hochreiter and J{\"u}rgen Schmidhuber.
\newblock Long short-term memory.
\newblock {\em Neural Computation}, 9(8):1735--1780, 1997.

\bibitem{shakespeare}
William Shakespeare.
\newblock The complete works of william shakespeare.

\bibitem{DBLP:journals/corr/abs-1812-01097}
Sebastian Caldas, Peter Wu, Tian Li, Jakub Konecn{\'{y}}, H.~Brendan McMahan,
  Virginia Smith, and Ameet Talwalkar.
\newblock {LEAF:} {A} benchmark for federated settings.
\newblock {\em CoRR}, abs/1812.01097, 2018.

\bibitem{tanner1987calculation}
Martin~A Tanner and Wing~Hung Wong.
\newblock The calculation of posterior distributions by data augmentation.
\newblock {\em Journal of the American Statistical Association},
  82(398):528--540, 1987.

\bibitem{van2001art}
David~A Van~Dyk and Xiao-Li Meng.
\newblock The art of data augmentation.
\newblock {\em Journal of Computational and Graphical Statistics}, 10(1):1--50,
  2001.

\bibitem{DBLP:journals/corr/IoffeS15}
Sergey Ioffe and Christian Szegedy.
\newblock Batch normalization: Accelerating deep network training by reducing
  internal covariate shift.
\newblock {\em CoRR}, abs/1502.03167, 2015.

\bibitem{DBLP:journals/corr/abs-1803-08375}
Abien~Fred Agarap.
\newblock Deep learning using rectified linear units (relu).
\newblock {\em CoRR}, abs/1803.08375, 2018.

\bibitem{zhu2020federated}
Hangyu Zhu, Haoyu Zhang, and Yaochu Jin.
\newblock From federated learning to federated neural architecture search: A
  survey.
\newblock {\em arXiv preprint arXiv:2009.05868}, 2020.

\bibitem{tensorflow2015-whitepaper}
Mart\'{\i}n Abadi, Ashish Agarwal, Paul Barham, Eugene Brevdo, Zhifeng Chen,
  Craig Citro, Greg~S. Corrado, Andy Davis, Jeffrey Dean, Matthieu Devin,
  Sanjay Ghemawat, Ian Goodfellow, Andrew Harp, Geoffrey Irving, Michael Isard,
  Yangqing Jia, Rafal Jozefowicz, Lukasz Kaiser, Manjunath Kudlur, Josh
  Levenberg, Dandelion Man\'{e}, Rajat Monga, Sherry Moore, Derek Murray, Chris
  Olah, Mike Schuster, Jonathon Shlens, Benoit Steiner, Ilya Sutskever, Kunal
  Talwar, Paul Tucker, Vincent Vanhoucke, Vijay Vasudevan, Fernanda Vi\'{e}gas,
  Oriol Vinyals, Pete Warden, Martin Wattenberg, Martin Wicke, Yuan Yu, and
  Xiaoqiang Zheng.
\newblock {TensorFlow}: Large-scale machine learning on heterogeneous systems,
  2015.
\newblock Software available from tensorflow.org.

\end{thebibliography}

%








\end{document}